\documentclass[12pt]{article}

\usepackage{amsmath,amssymb,amsbsy,amsfonts,latexsym,graphicx}
\usepackage{hyperref}
\usepackage{color, array, subfigure}
\usepackage{cite}
\usepackage[title]{appendix}

\allowdisplaybreaks

\evensidemargin \oddsidemargin

\begin{document}
%%%%%%%%%%%%%%%%%%%%%%%%%%% TITLE PAGE %%%%%%%%%%%%%%%%%%%%%%%%%%%%%%%
\begin{titlepage}

%-------------------- footnote symbol in title page -----------------
\renewcommand{\thefootnote}{\fnsymbol{footnote}}

%----------------------- preprint number & date ---------------------

%\begin{flushright}
%???-??????
%\end{flushright}

%---------------------------- Title ---------------------------------
\vspace{15mm}
\baselineskip 9mm
\begin{center}
  {\Large \bf 
{{D-instanton Effects on the Holographic Weyl Semimetals}} 
 }
\end{center}

%--------------------- Authors and Addresses ------------------------
\baselineskip 6mm
\vspace{10mm}
\begin{center}
Hwajin Eom$^{a}$,  and Yunseok Seo$^{b}$
 \\[10mm] 
 {${}^a$\sl Dasan University College, Ajou University, Gyunggi-do 16499, Korea}\\
 {${}^b$\sl College of General Education, Kookmin University, Seoul 02707, Korea}
   \\[3mm]
  {\tt  ${}^a$eom16@ajou.ac.kr, ${}^b$yseo@kookmin.ac.kr}
\end{center}

\thispagestyle{empty}

%-------------------------- abstract --------------------------------
%\vfill
\vspace{1cm}
\begin{center}
{\bf Abstract}
\end{center}
\noindent
We investigate the effects of D-instantons on a holographic Weyl semimetal in a top-down approach. We find that the presence of D-instantons induces a novel phase transition from a Weyl semimetal to a gapped phase, providing a new mechanism for gap formation in strongly coupled systems.
By analyzing the free energy of probe D7-brane embeddings, we construct the phase diagram in terms of the fermion mass, instanton density, and temperature, measured in units of the Weyl parameter. We show that, in addition to the conventional mass-driven transition associated with band inversion, the instanton density gives rise to a qualitatively different transition whose mechanism cannot be understood within the standard Dirac picture.
We compute nonlinear electric conductivities from the regularity condition at the black hole horizon and demonstrate that the two types of transitions exhibit distinct transport behavior. In particular, the instanton-induced transition leads to a gapped phase with properties that differ from the conventional insulating phase.
We also discuss possible interpretations of this phase in the boundary theory and comment on its relation to nonperturbative effects in strongly interacting systems.
\\ [15mm]
Keywords: Gauge/gravity duality, D-instanton, Weyl semimetal
%\\ PACS numbers:   ????

\vspace{5mm}
\end{titlepage}

%%%%%%%%%%%%%%%%%%%%%%%%%%%%%%%%%%
\section{Introduction}
%%%%%%%%%%%%%%%%%%%%%%%%%%%%%%%%%%
Weyl semimetals have been extensively studied in recent years due to their topological as well as electronic properties. In these systems, pairs of Weyl nodes appear in momentum space as a consequence of band inversion. The low-energy excitations near each nodal point are described by Weyl fermions \cite{Burkov:2011hve,hasan2017discovery,Yan:2016euz,Burkov:2017rgl,RevModPhys.90.015001,gao2019topological,li2023emergence}.
In a Weyl semimetal, each pair of Weyl nodes carries opposite chirality, which arises from the breaking of either spatial inversion or time-reversal symmetry. Each Weyl node is characterized by a nonzero Chern number with opposite sign, reflecting its chirality. These nodes act as monopoles of Berry curvature, leading to the emergence of topologically protected surface states known as \textit{Fermi arcs}.
Weyl semimetals have been experimentally observed in materials such as TaAs \cite{PhysRevX.5.031013}, NbAs \cite{xu2015discovery}, TaP \cite{xu2015experimental}, and related compounds \cite{PEI2021100509,gooth2019axionic}, through the observation of Fermi arcs and linear dispersion relations. More recently, transitions between topological insulators and Weyl semimetals have been reported in Cr-doped ${\rm Bi_2 Te_3}$ \cite{belopolski2025synthesis} and in UNiSn \cite{ivanov2019topological}, where strong spin-orbit coupling and time-reversal symmetry breaking play essential roles.
The simplest field-theoretic description of a Weyl semimetal is given by a massive Dirac fermion coupled to a non-dynamical axial gauge field $A_j^5$ in (3+1) dimensions \cite{colladay1998lorentz,grushin2012consequences}:
\begin{align}\label{WeylL}
{\cal L}= \bar{\psi} \left(i \gamma^{\mu}\partial_{\mu} - m + A_{j}^5 \gamma^{j} \gamma^5 \right) \psi,
\end{align}
where $\mu =0,1,2,3$ denotes spacetime indices and  $j=1,2,3$ denotes spatial indices. The axial gauge field $A_j^5$ breaks time-reversal symmetry. For a simple choice of $A_j^5 = \frac{b}{2} \delta_{jz}$, the energy eigenvalues are given by

\begin{align}\label{energyE}
\epsilon= \pm \sqrt{k_x^2+k_y^2 +\left(\frac{b}{2}\pm \sqrt{k_z^2+m^2} \right)^2}.
\end{align}

From the dispersion relation, two of the four branches (those with the positive sign inside the parentheses) remain gapped for finite $m$ and $b$. In the strong axial field regime, $|m/b|<1/2$, two gapless points appear at $\vec{k} =(0,0,\pm\sqrt{(b/2)^2 -m^2})$,
which correspond to Weyl nodes. Near each node, the dispersion relation becomes linear, characteristic of Weyl fermions. In contrast, for $|m/b|>1/2$, a finite energy gap opens and the system becomes an insulator. At the critical point 
$|m/b|=1/2$, the two nodes merge at the origin of momentum space, signaling a band inversion as the parameter $b$ is varied.

The axial vector potential in (\ref{WeylL}) gives rise to an axial anomaly that leads to nontrivial transport phenomena. In particular, an electric current is generated orthogonal to both the applied electric field and the direction of the axial vector, resulting in an anomalous Hall conductivity
\begin{align}
\sigma_{xy} = \frac{1}{4\pi} \sqrt{b^2 -4m^2} ,\Theta (|b| - 2|m|),
\end{align}
where $\Theta$ is the Heaviside step function. For further reviews on Weyl semimetals, see \cite{rao2016weyl,wang2017quantum,zhong2025weyl}.

The main purpose of this paper is to investigate strong interaction effects in Weyl semimetals using gauge/gravity duality. One may ask why strong interactions should be relevant if the system is described by a free Dirac theory. The key point is that the free Dirac description is valid when the Fermi surface is large. However, when the Fermi energy approaches the Dirac point, the one-particle description breaks down and interaction effects become important.
In systems with a small Fermi surface, screening of the Coulomb interaction becomes inefficient, leading to strong correlations. For instance, graphene is well described by a free Dirac theory near a large Fermi surface, where electrons behave as weakly interacting quasiparticles. However, near the Dirac point, deviations such as the violation of the Wiedemann--Franz law are observed \cite{crossno2016observation}. Similarly, in magnetically doped topological insulators, a transition from weak anti-localization (WAL) to weak localization (WL) has been observed \cite{PhysRevLett.108.036805,zhang2012interplay,bao2013quantum}, which cannot be explained by the Hikami--Larkin--Nagaoka formula \cite{10.1143/PTP.63.707}.

Gauge/gravity duality \cite{maldacena1999large,witten1998anti,gubser1998gauge} provides a powerful framework to study strongly interacting systems. In particular, it allows one to compute DC and AC transport coefficients in strongly coupled systems \cite{andrade2014simple,blake2013universal,donos2014thermoelectric,kim2014coherent,seo2016character}. Holographic models have also successfully reproduced experimental features such as transport with multiple currents \cite{seo2017holography} and magnetoelectric responses in topological insulators \cite{seo2017strong,seo2017small}.

There are two main classes of holographic approaches to Weyl semimetals. The first is the \textit{bottom-up} approach, based on a five-dimensional Einstein--Maxwell theory with two $U(1)$ gauge fields, where one of them plays the role of an axial gauge field through Chern--Simons interactions \cite{Landsteiner:2015lsa,Landsteiner:2015pdh,Liu:2018spp,Landsteiner:2019kxb,Juricic:2020sgg}. These models have been used to study phase transitions and anomalous Hall conductivities. Various works have also investigated surface states and fermionic spectral functions in this framework \cite{Ammon:2016mwa,Landsteiner:2016stv,Baggioli:2018afg,Grignani:2016wyz,Ammon:2018wzb,Liu:2018djq,Baggioli:2020cld,Oh:2020cym}.

The second class is the \textit{top-down} approach, based on type IIB string theory on $AdS_5 \times S^5$, corresponding to the large-$N_C$ limit of D3-branes. The dual boundary theory is a four-dimensional ${\cal N}=4$ $SU(N_C)$ supersymmetric Yang--Mills theory with $SO(6)$ R-symmetry. Introducing $N_F$ probe D7-branes adds ${\cal N}=2$ hypermultiplets in the fundamental representation. The axial symmetry arises from the breaking $SO(6) \rightarrow SO(4)\times U(1)_A$.
In \cite{Hashimoto:2016ize,Kinoshita:2017uch}, axial gauge fields were generated via rotating electric fields in the $x-y$ plane, and non-equilibrium steady states were studied.

More recently, an axial current has been introduced by imposing periodicity of the probe D7-brane along a transverse direction \cite{BitaghsirFadafan:2020lkh}. By analyzing the D7-brane embedding, a phase transition between a Weyl semimetal and a trivial insulator was identified in terms of the fermion mass $m_e$ and the Weyl parameter $b$. Based on this setup, AC conductivities and fermionic spectral functions have been investigated in \cite{Furukawa:2024zet,Lu:2025zxq}.
Since the free energy depends on the worldvolume of the probe D7-brane, the phase structure is sensitive to the background geometry. In previous studies, it was shown that D-instantons can significantly affect D7-brane embeddings \cite{Gwak:2012ht}. In this work, we study the effect of D-instantons on the holographic Weyl semimetal model of \cite{BitaghsirFadafan:2020lkh}. We analyze the phase structure in terms of the Weyl parameter and the instanton density, and we also investigate nonlinear electric conductivities following \cite{Karch:2007pd,OBannon:2007cex}.

This paper is organized as follows. In section 2, we study the phase structure of the Weyl semimetal using probe D7-branes in the D3/D-instanton background. In section 3, we compute nonlinear conductivities from the regularity condition of the probe D7-brane and investigate anomalous Hall transport. In section 4, we conclude and discuss the physical interpretation of bulk parameters in the boundary theory.

\section{Weyl semimetal to insulator transition}
\subsection{Holographic model for Weyl semimetal}

In this section, we briefly review the top-down holographic model for a Weyl semimetal proposed in \cite{BitaghsirFadafan:2020lkh}, as well as its extension including D-instantons. In type IIB supergravity in ten-dimensional spacetime, we consider a system of $N_C$ coincident D3-branes and $N_F$ coincident D7-branes. The  ${\cal N}=2$ supersymmetric configuration of this system is summarized in Table \ref{table:D3D7}. On the D3-brane worldvolume, ${\cal N}=4$ super Yang--Mills theory is realized at low energies, consisting of a gauge field, four Weyl fermions, and six scalar fields. The theory has an $SO(6)_R$ symmetry corresponding to rotations in the directions transverse to the D3-branes.

\begin{table}
    \centering
    \begin{tabular}{|c|cccccccccc|}
    \hline
          & $x_0$ & $x_1$ & $x_2$ & $x_3$ & $x_4$ &  $x_5$ &  $x_6$ &  $x_7$ & $x_8$& $x_9$ \\
    \hline      
        D3 & $\times$ & $\times$ &$\times$  & $\times$ &  &  &  &  & & \\
        D7 &  $\times$ & $\times$ & $\times$ & $\times$ &$\times$  &$\times$  & $\times$ &$\times$  & &\\
    \hline
    \end{tabular}
    \caption{Brane configuration of the $N_C$ D3 branes and $N_F$ D7 branes preserving ${\cal N}=2$ SUSY.}
    \label{table:D3D7}
\end{table}

Open strings stretching between the D3- and D7-branes give rise to $N_F$ ${\cal N}=2$ hypermultiplets in the fundamental representation of the $SU(N_C)$ gauge group. These fields can be interpreted as `quarks' or `electrons' in the boundary theory. In this paper, we refer to this Dirac fermion as an `electron', carrying a $U(1)$ charge with mass $m_e$, which is determined by the separation between the D3 and D7-branes in the $(x_8,x_9)$ directions \cite{Chesler:2006gr,Kirsch:2006he,Erdmenger:2007cm}.

We parametrize the transverse directions to the D7-brane as
\begin{equation}
W(x,r) = x_8 + i x_9,
\end{equation}
and consider the embedding
\begin{equation}
W(x,r) = \rho(r) e^{i\phi(x)}
\end{equation}
where $\rho(r)$ determines the radial profile (related to the fermion mass) and $\phi(x)$ is a spacetime-dependent phase, interpreted as a pseudoscalar field.

The D7-brane fermion action (after $\kappa$-symmetry fixing) takes the schematic form
\begin{equation}
S \sim \int d^8\xi \sqrt{-g}
\bar{\Psi}\left( \gamma^{\mu}D_{\mu} +m_e \cdots \right)\Psi,
\end{equation}
where $m_e$ is a fermion mass which is a separation between D3 and D7-branes.
Due to the curved embedding, fermions couple to transverse rotations through the spin connection, which induces the term
\begin{equation}
\bar{\Psi}\left( \frac{i}{2} \gamma^{\mu} \partial_{\mu} \phi \Gamma^{89} \right)\Psi
\end{equation}
where $\Gamma^{89}$ generates rotations in the $(x_8,x_9)$ plane.
Using gamma matrix decomposition together with $\kappa$-symmetry projection, one finds
\begin{equation}
\Gamma^{89} \longrightarrow \pm \gamma_5.
\end{equation}
Thus, the effective coupling becomes
\begin{equation}
\bar{\psi} \gamma^\mu (\partial_\mu \phi) \gamma_5 \psi.
\end{equation}
Combining this with the kinetic and mass terms, the four-dimensional effective Lagrangian for fermionic fluctuations on the D7-brane is given by
\begin{equation}
\label{eq:action_Dirac}
\mathcal{L}_{\psi} = i\bar{\psi}
\left(i\gamma^\mu \partial_\mu -m_e
+\frac{\partial_\mu \phi}{2} \gamma^\mu \gamma^5 \right) \psi,
\end{equation}
where $\mu = 0,1,2.3$ and $\phi$ is a angle direction of polar coordinates in $(x_8,x_9)$ plane. For a linear profile $\phi = b z$, the 
effective Lagrangian (\ref{eq:action_Dirac}) reduced to (\ref{WeylL}) with $A_j^5 = b/2 \delta_{jz}$. Geometrically, this corresponds to D7-branes spiraling around the D3-branes in the $(x_8,x_9)$ plane as they extend along $z$ direction. The parameter $b$ plays the role of the Weyl node separation in the boundary theory.

Based on this D3/D7 setup, it was shown in \cite{BitaghsirFadafan:2020lkh} that there exists a phase transition between a Weyl semimetal phase and a trivial insulating phase, controlled by the temperature and the Weyl parameter $b$. We will review these results in the next section.

n this work, we are interested in nonperturbative topological effects in the Weyl semimetal. A natural candidate for such effects is provided by instantons. In the holographic framework, instantons are described by D-instantons in type IIB supergravity. The background solution corresponding to a uniform distribution of D-instantons on D3-branes was constructed in \cite{Liu:1999fc}, and its finite-temperature extension was developed in \cite{Ghoroku:2005tf}.

We introduce probe D7-branes into this D3/D-instanton background, where the dilaton field is nontrivial and encodes the instanton density. We consider embeddings with a nontrivial angular profile $\phi=b z$, corresponding to a finite Weyl parameter. We then study how the instanton density and the Weyl parameter affect the D7-brane embeddings and the resulting phase structure.

\subsection{Background geometry and D7 brane embeddings}
n this section, we use a finite-temperature extension of the D3/D-instanton background. To obtain an analytic solution, the D-instantons are uniformly smeared over the D3-brane worldvolume. The ten-dimensional supergravity action in the Einstein frame is given by
\begin{equation}
\label{eq:action_Eframe}
S=\frac{1}{\kappa} \int d^{10} x \sqrt{-g}
\left(
R-\frac{1}{2} \left(\partial\Phi \right)^2
+\frac{1}{2} e^{2\Phi} \left(\partial\chi \right)^2
-\frac{1}{6} F^2_{(5)}
%\textcolor{red}{+\mathcal{L}_\psi}
\right).
\end{equation}
where $\Phi$ and $\chi$ denote the dilaton and the axion, respectively, and $F_{(5)}$ is a five-form field strength of the Ramon-Ramon field sourced by D3 branes. The dilaton term in  (\ref{eq:action_Eframe}) can be canceled by choosing the axion term as $\chi =-e^{-\Phi} +\chi_0$, then the solutions in string frame become
\begin{align}
\label{eq:metric}
& ds^2_{10} =e^{\Phi/2}
\left[
\frac{r^2}{L^2} \left(
-f(r)^2 dt^2 +d\vec{x}^2 \right)
+\frac{1}{f(r)^2} \frac{L^2}{r^2} dr^2
+L^2 d\Omega^2_5
\right],\\\label{eq:sol_Sframe}
& e^\Phi = 1+\frac{q}{r_H^4} \log\frac{1}{f(r)^2},
\quad \chi = -e^{-\Phi} +\chi_0,
\quad f(r) = \sqrt{1-\bigg( \frac{r_H}{r}\bigg)^4},
\end{align}
where $\vec{x} =\left(x,y,z\right)$ and $L^4 = 4\pi g_s N_c \alpha'^2$. The constant $q$ is the D-instanton charge, representing the instanton density.

We introduce a dimensionless coordinate $\xi$ 
defined by
$\frac{d\xi^2}{\xi^2} =
\frac{dr^2}{r^2 f(r)^2}$.
Then the background geometry \eqref{eq:metric} becomes
\begin{equation}
\label{eq:metric_xi}
ds^2_{10} =e^{\Phi/2}
\left[
\frac{r^2}{L^2} \left(
-f(r)^2 dt^2 +d\vec{x}^2 \right)
+\frac{L^2}{\xi^2}
\left(
d\xi^2+\xi^2 d\Omega^2_5
\right)
\right].
\end{equation}
The relations between $r$ and $\xi$ 
can be written as
\begin{equation}
\label{eq:rel_radial}
\bigg(\frac{r}{r_H}\bigg)^2 =\frac{1}{2} \left(\frac{\xi^2}{\xi_H^2}+\frac{\xi_H^2}{\xi^2}\right),
\quad
f(r) = \left(\frac{1-\xi^4_H /\xi^4}{1+\xi^4_H /\xi^4}\right) \equiv \frac{\omega_-}{\omega_+},
\quad
\omega_\pm \equiv 1\pm \frac{\xi^4_H}{\xi^4}.
\end{equation}
The condition that $r$ and $\xi$ coincide at the asymptotic region gives the relation $r_H = \sqrt{2} \xi_H$.

To describe the embedding of the probe D7-brane,
we decompose $\mathbb{R}^6$ part in \eqref{eq:metric_xi}
into $\mathbb{R}^4 \times\mathbb{R}^2$ such as
\begin{align}
\label{eq:metric_decomposed}
ds^2_{10} & =e^{\Phi/2}
\left[
\frac{r^2}{L^2} 
\left(
-\frac{\omega^2_-}{\omega^2_+} dt^2
+d\vec{x}^2 \right)
+\frac{L^2}{\xi^2}
\left(
d\rho^2+\rho^2 d\Omega^2_3
+dR^2 + R^2 d\phi^2
\right)
\right], \\
\label{eq:C4}
C_4 & = 
%\textcolor{red}{e^\Phi}
\frac{\xi^4}{L^4} \omega_+^2
dt\wedge dx\wedge dy\wedge dz
-
%\textcolor{red}{e^{\Phi/2}}
\frac{L^4 \rho^4}{\xi^4} d\phi\wedge d\Omega_3,
\end{align} 
where $\xi =\sqrt{\rho^2 + R^2}$
and $\omega (S^3)$ denotes the volume form on a unit-radius $S^3$.
D7-brane extends along $\left(r,\vec{x},\rho\right)$, wraps $S^3$,
and is transverse to $R$ and $\phi$ direction.

%For the embedding $R=R(\rho)$,
%the induced metric on D7 brane 
%is given as
%\begin{equation}
%\label{eq:metric_D7}
%ds^2_{D7} =e^{\Phi/2}
%\left[
%\frac{r^2}{L^2} 
%\left(
%-\frac{\omega^2_-}{\omega^2_+} dt^2
%\!+\! d\vec{x}^2 \right)
%+\frac{L^2}{\xi^2}
%\!\!\left(\!\!
%\!\left(\!1+\left(\frac{\partial R}{\partial \rho}\right)^2 \!\right)^2\!\! d\rho^2
%+\rho^2 d\Omega^2_3 + R^2 \left(\frac{\partial \phi}{\partial z} \!\right)^2dz^2\!
%\!\right)\!
%\right],
%\end{equation}
%where $R'$ denotes
%$\partial R(\rho) /\partial\rho$.

The D7-brane action consists of an abelian Dirac-Born-Infeld (DBI) action, the Wess-Zumino (WZ) term, and Chern-Simons
terms,
%Chern-Simons (CS)
\begin{align}
\notag
S_{D7} & = S_{DBI} +S_{WZ} 
+S_{CS}
\\\label{eq:D7_action}
& =-N_f \mu_7 \int d^8\sigma e^{-\Phi}
\sqrt{-{\rm det} \left(P[g]+2\pi\alpha' F\right)}
+2\pi^2 \alpha'^2 N_f \mu_7 \int P[C_4] \wedge F \wedge F 
+S_{CS},
\end{align}
where $\mu_7$ is the D7-brane tension,
$F=dA$ is the field strength for a $U(1)$ world-volume gauge field $A$,
and $P[g]$ and $P[C_4]$
are
the pullback of the metric and four-form in \eqref{eq:C4}.

To construct embeddings dual to a Weyl semimetal, we take
\begin{equation}
\label{eq:WSM_setting}
A=0,
\qquad R=R(\rho) \qquad
\phi=bz.
\end{equation}
With the ansatz \eqref{eq:WSM_setting}, the WZ term vanishes. The induced metric becomes
\begin{align}
\label{eq:metric_WSM}
ds^2_{D7} =
&e^{\Phi/2}
\Big[
\frac{r^2}{L^2} 
\left(
-\frac{\omega^2_-}{\omega^2_+} dt^2
+\left(dx^2 +dy^2 \right) \right)
+\left(\frac{r^2}{L^2} +\frac{L^2 b^2 R^2}{\xi^2} \right) dz^2 \cr
&+\frac{L^2}{\xi^2}
\left(
\left(1+R'^2 \right)d\rho^2
+\rho^2 d\Omega^2_3 
\right)
\Big],
\end{align} 
where $R'=\partial R(\rho) /\partial\rho$.
Then DBI action for D7-brane is given by\footnote{We consider only DBI action, since the Chern-Simons term does not contribute to the equation of motion.}
\begin{equation}\label{eq:D7_action_potential}
S_{D7} \equiv -\tau_7 \int dt d\rho ~ V(\rho,R) \sqrt{1+R'^2},
\end{equation}
where
\begin{equation}
\label{eq:potential} 
V(\rho,R) = e^{\Phi} \omega_{+}\omega_{-}
\rho^3 \sqrt{1+ \frac{L^4 b^2 R^2}{\omega_{+}\xi^4}},\quad
\tau_7 = 2\pi^2 N_f \mu_7  .
\end{equation}
The equation of motion for the DBI action can be written as
\begin{equation}
\label{eq:D7_EOM}
\frac{R''}{1+R'^2} +
R' \frac{\partial \log V}{\partial \rho}
-\frac{\partial \log V}{\partial R} =0.
\end{equation}

There are two types of embeddings: the black hole embedding, where the D7-brane reaches the horizon, and the Minkowski embedding, where it does not. For the Minkowski embedding, regularity at $\rho=0$ requires
\begin{align}\label{MEbc}
R(\rho=0) = R_0,~~~R'(\rho=0) =0.
\end{align}
For the black hole embedding, regularity at the horizon $\rho=\rho_H$ requires
\begin{align}\label{BEbc}
R(\rho_H) = R_H =\sqrt{\xi_H^2 - \rho_H^2},~~~R'(\rho_H) = \frac{R_H}{\rho_H},
\end{align}
which implies that the D7-brane touches the horizon in the radial direction.

In the asymptotic region,  $e^\Phi$ and $\omega_\pm$ become one and $\xi \rightarrow \rho$, then the equation of motion (\ref{eq:D7_EOM}) reduces to
\begin{equation}
\label{eq:asym_D7_EOM}
R''+\frac{3R'}{\rho} 
-\frac{L^4 b^2  R}{ \rho^4} =0,
\end{equation}
where we assume that the D7 brane does not change much in the asymptotic region($R' \ll 1$).
For $\tilde{b} \equiv L^2 b$,
we get the asymptotic form of $R(\rho)$.
\begin{equation}
R(\rho) = \frac{c_1}{2\sqrt{\pi} \tilde{b}^2}
+\frac{-c_1 +2 c_2 \gamma + 8 \sqrt{\pi}c_2 \log\big(\frac{\tilde{b}}{2\rho}\big)}{8\sqrt{\pi}\rho^2} +
O\bigg(\frac{\log\rho}{\rho^4}\bigg),
\end{equation}
where $\gamma$ is Euler's constant
and $c_1,~c_2$ are integration constants.
By using the following relations, such as
$c_1=2\sqrt{\pi}\tilde{b}^2 m_e$
and
$c_2=c-\tilde{b}^2 m_e \left(\frac{2\gamma-1}{4}+\frac{1}{2}\log \big(\frac{\tilde{b}}{2L}\big) \right)$,
we get
\begin{equation}
\label{eq:asym_D7}
R(\rho) =
m_e
\left(1-\frac{\tilde{b}^2}{2\rho^2} \log \left(\frac{\rho}{L}\right)
\right)
+\frac{c}{\rho^2}
+ O\bigg(\frac{\log\big(\frac{\rho}{L}\big)}{\rho^4}\bigg),
\end{equation}
where the integration constants $m_e$ and $c$ are interpreted as the fermion mass and the chiral condensation, respectively. This asymptotic form is the same as \cite{BitaghsirFadafan:2020lkh} because the D-instanton effect is suppressed at infinity. 

Holographically, $m_e$ corresponds to the separation between D3 and D7-branes and hence to the fermion mass. Although these degrees of freedom originate from fundamental matter in the gauge theory, we refer to them as `electrons' to emphasize their role in the effective condensed matter description.

\begin{figure}[ht]
\begin{center}
\subfigure[$q=0,~m_e =1.5$]{\includegraphics[width=0.4\textwidth]{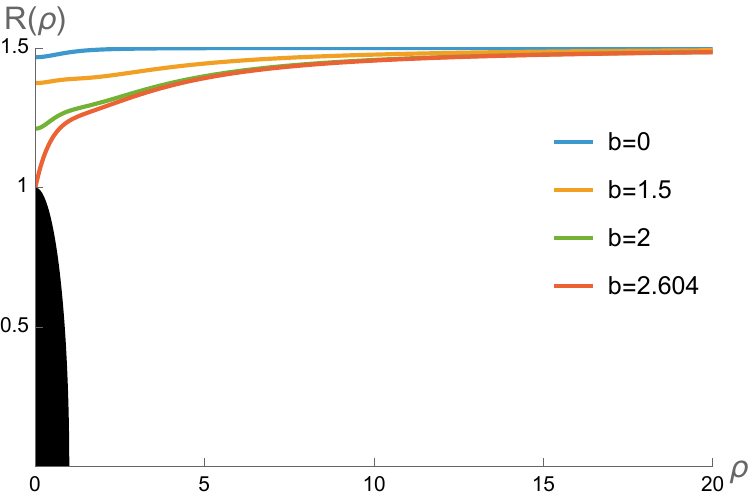}}
\hskip.5cm
\subfigure[$q=0,~m_e =0.7$]{\includegraphics[width=0.4\textwidth]{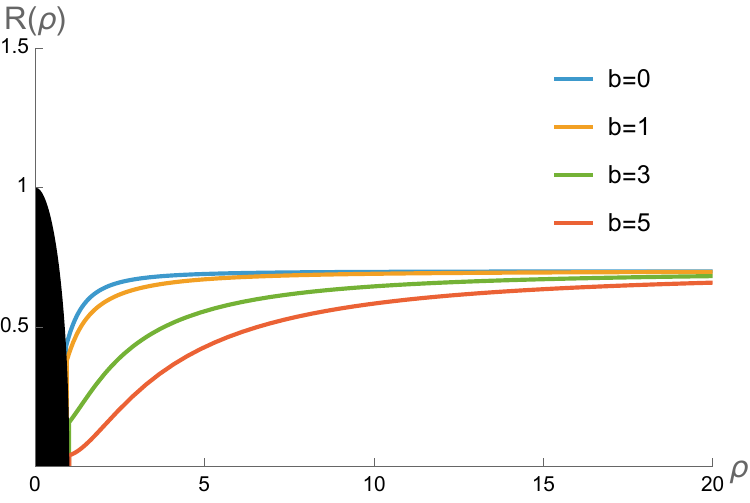}}
\subfigure[$b=0,~m_e =1.5$]{\includegraphics[width=0.4\textwidth]{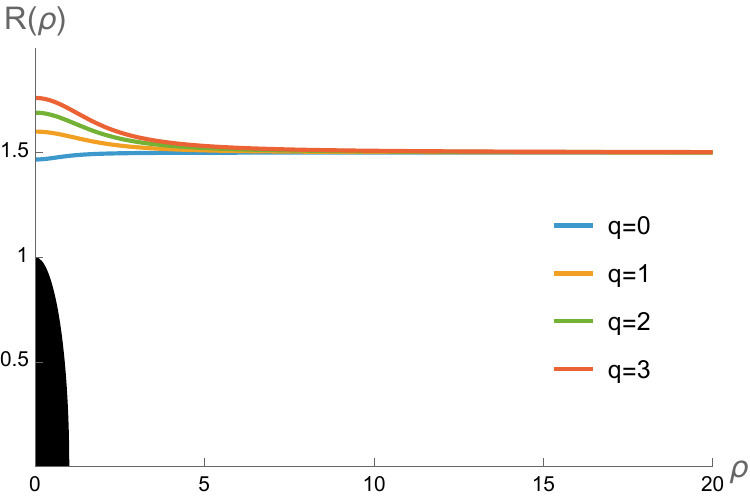}}
\hskip.5cm
\subfigure[$b=0,~m_e =0.7$]{\includegraphics[width=0.4\textwidth]{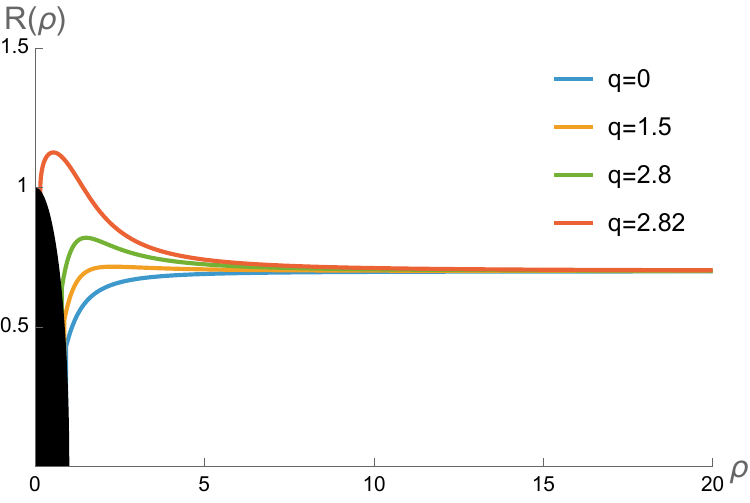}}
\caption{D7 brane embeddings without $q$ (a), (b) and without $b$ (c), (d). Here, we set $\xi_h=1$, which is shown as a black disc.}
\label{fig:D7Embs}
\end{center}
\end{figure}

With appropriate boundary conditions (\ref{MEbc}) or (\ref{BEbc}), we can solve the equation of motion for the D7-brane numerically. The embedding solutions are shown in Figure \ref{fig:D7Embs}.

To illustrate the effect of the Weyl parameter $b$ and the D-instanton density $q$, we set $q=0$ in Figure \ref{fig:D7Embs} (a), (b) and $b=0$ in Figure \ref{fig:D7Embs} (c), (d). In the absence of the D-instantons, the D7 brane embeddings agree with \cite{BitaghsirFadafan:2020lkh}. For fixed $m_e$, increasing $b$ pulls the D7-brane toward the black hole horizon both in the Minkowski embedding and the black hole embedding. In the Minkowski embedding, the brane bends downward and eventually touches the horizon at a critical value $b=b_*$. For $b>b_*$, only black hole embeddings exist.

In contrast, the D-instanton density pushes the D7-brane away from the horizon. As shown in Figure \ref{fig:D7Embs} (c), (d), increasing $q$ introduces a repulsive effect. For sufficiently large $q>q_*$, black hole embeddings no longer exist. Therefore, the phase diagram is expected to be dominated by black hole embeddings at small $b$ and $q$, while large $b$ or large $q$ favor Minkowski embeddings.

\subsection{Zero temperature analysis}
In this section, we perform a zero-temperature analysis of the probe D7-brane. When $T=0$, the background geometry reduces to pure $AdS_5 \times S^5$ and hence $\xi_H =0$ and $\omega_\pm \rightarrow 1$.
The potential \eqref{eq:potential} then simplifies to
\begin{equation}
\label{eq:potential_T0}
V_{T=0} (\rho,R) = 
\left(1+\frac{q}{\left(\rho^2 +R^2 \right)^2 } \right)
\rho^3
\sqrt{1+ \frac{L^4 b^2 R^2}{\left(\rho^2 +R^2 \right)^2}}
\end{equation}
In addition,
the equation of motion \eqref{eq:D7_EOM}
becomes
\begin{equation}
\label{eq:D7_EOM_T0}
\frac{R''}{1+R'^2} +
\frac{R \left(-b^2 L^4 \left(\rho ^6 + q \rho ^2\right)+4 q \rho ^4
+\mathcal{O} (R^2)\right)
+R' \left(3 \rho ^9-q \rho ^5 +\mathcal{O} (R^2)\right)}{\rho^6 \left(q+\rho^4 \right) \left(1+\frac{L^4 b^2 R^2}{\rho^4}\right)}=0.
\end{equation}
Even at zero temperature, there exist two types of embeddings, distinguished by whether 
$R(\rho =0 )=0$ or not.
The  embedding with $R(0 )  \neq 0$
corresponds to the Minkowski embedding, with the same initial conditions as in the finite-temperature case \eqref{MEbc}.

Expanding $R(\rho)$ near $\rho=0$ in \eqref{eq:D7_EOM_T0},
we obtain 
\begin{equation}
  \label{eq:D7_EOM_T0_Mink}
\frac{R''}{1+R'^2} = \frac{3 R'}{\rho} +\mathcal{A},
\quad \mathcal{A} = \frac{b^2 L^4 R_0^4 + q\left(5 b^2 L^4 +4  R_0^2 \right)}{R_0  \left(b^2 L^4+R_0^2\right) \left(R_0^4 +q\right)}
\end{equation}
where $\mathcal{A}$ is just a constant.
The solution of \eqref{eq:D7_EOM_T0_Mink}
for a small $\rho$
with the initial condition \eqref{MEbc} is
\begin{equation}
    \label{eq:eq:D7_sol_T0_Mink}
    R\left(\rho\right) = R_0 -\frac{\mathcal{A}}{8} \rho^2 +\mathcal{O} \left(\rho^4\right).
\end{equation}
Note that $e^\Phi$ goes to $1$ in the asymptotic region
and $\omega_\pm=1$ in the zero temperature limit.
Therefore, the near-boundary asymptotic expansion of $R(\rho) $ remains 
\begin{equation}
R(\rho)=m_e +\mathcal{O} \left(\frac{\log \rho}{\rho^2} \right)
\end{equation}
as in the finite-temperature case.

The other type of embedding with $R(0) =0$
corresponds to the black hole embedding
and should be regular as $\rho\rightarrow 0$.
From \eqref{eq:D7_EOM_T0}
the linearized equation can be obtained as
\begin{equation}
\label{eq:D7_EOM_T0_lin}
R''+ \frac{
R\left(-b^2 L^4 \left(\rho ^4+q\right)+4 q \rho ^2 \right)
+R' \left(3 \rho ^7-q \rho ^3\right) 
}{\rho ^4 \left(\rho ^4+q\right)} =0 .
\end{equation}
In order to find a regular solution of \eqref{eq:D7_EOM_T0_lin},
we consider the following form
\begin{equation}
\label{eq:T0_BH_form}
R(\rho) \equiv \frac{\rho^2}{\sqrt{q+\rho^4}} \mathcal{R} (\rho),
\end{equation}
then the linearized equation \eqref{eq:D7_EOM_T0_lin}  for a small $\rho$ is rewritten as
\begin{equation}
\label{eq:D7_EOM_T0_lin_sim_smallrho}
-\frac{b^2 L^4 \mathcal{R}}{\rho ^4}+\mathcal{R}''+\frac{3 \mathcal{R}'}{\rho } =0 .
\end{equation}
Since $\mathcal{R}(\rho)\to R(\rho) $ in the asymptotic region, the equation \eqref{eq:D7_EOM_T0_lin_sim_smallrho} reduces to the same form in the nonzero temperature case \eqref{eq:asym_D7_EOM}
and then the asymptotic behavior of $R(\rho) $ also remains the same.
The regular solution of \eqref{eq:D7_EOM_T0_lin_sim_smallrho} is given by
\begin{equation}
\label{eq:smallrho_sol}
R(\rho) \approx \frac{\rho^2}{\sqrt{q+\rho^4}} \frac{L^2 b\phantom{`} \eta}{\rho} K_1 \left(\frac{L^2 b}{\rho} \right)
=\frac{L^2 b\phantom{`} \eta \rho}{\sqrt{q+\rho^4}} K_1 \left(\frac{L^2 b}{\rho} \right),
\end{equation}
where $\eta$ is an integration constant and $K_1 (x)$ is the modified Bessel function and it  vanishes exponentially as $x \to\infty$.
Therefore, the solution for a small $\rho$ is
\begin{equation}
    \label{eq:zero_bh_embb}
R (\rho) = \sqrt{\frac{\pi L^2 b}{2}} \eta 
\frac{e^{-\frac{b L^2}{\rho }}}{\sqrt{\rho}}
\left[1 -\frac{q \left(1+\mathcal{O} (\rho )\right)}{2 \rho ^{4}} +\mathcal{O} (\rho^2) \right],
\end{equation}
which indicates that it corresponds to the black hole embedding
at zero temperature limit.

\begin{figure}[ht]
\begin{center}
\subfigure[Minkowski embeddings]{\includegraphics[width=0.4\textwidth]{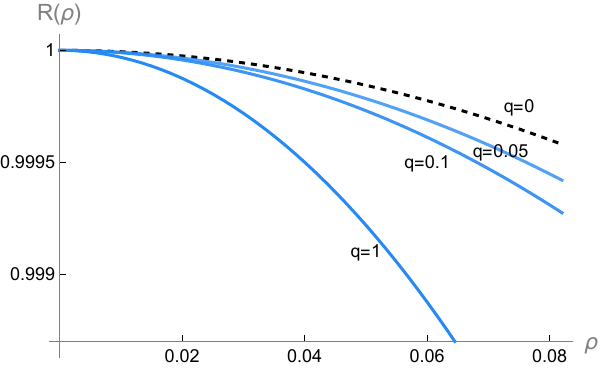}}
\hskip.5cm
\subfigure[Black hole embeddings]{\includegraphics[width=0.4\textwidth]{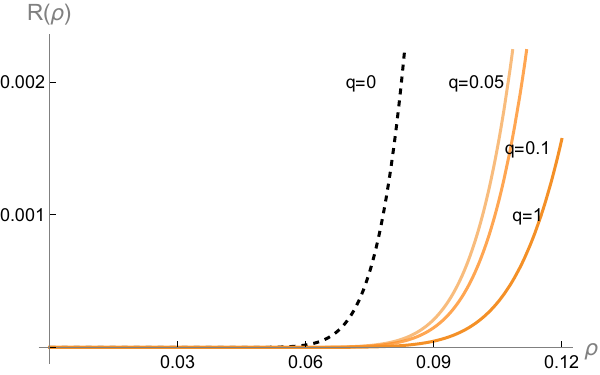}}
\caption{$q$ dependence of D7 brane embeddings near $\rho=0$ with $T=0$. Here, we set $b=1$ and $L=R_0 =1$ for simplicity. }
\label{fig:T0Embs}
\end{center}
\end{figure}

The D-instanton density dependence of each embedding is shown in Figure \ref{fig:T0Embs}. In the figure, we fix the position of D7-branes at $\rho=0$, instead of fixing $m_e$ for simplicity. Because of it, the embedding behavior as $q$ increases seems to behave oppositely to Figure \ref{fig:D7Embs} (c) and (d). But, one can easily notice that their behaviors agree with each other.

There is one comment on the embeddings near $\rho=0$ we derived.
In the limit of $\rho\to 0$, one can easily check that
the Minkowski embedding \eqref{eq:eq:D7_sol_T0_Mink}
and the black hole embedding  \eqref{eq:zero_bh_embb}
smoothly reduce to the known results in the limit $q\to 0$. In this limit,
\begin{equation}
    \mathcal{A} \to \frac{b^2 L^4}{R_0 \left(b^2 L^4 +R_0^2\right)},
    \qquad
    \mathcal{R}(\rho)\to R(\rho) .
\end{equation}
Then the solution becomes
%\eqref{eq:eq:D7_sol_T0_Mink} and \eqref{eq:zero_bh_embb}
%become
\begin{equation}
    \label{eq:zero_Mink_embb_q0}
    R(\rho) = R_0 -\frac{b^2 L^4}{8R_0 \left(b^2 L^4 +R_0^2\right)} \rho^2 +\mathcal{O} (\rho^4 ) , 
\end{equation}
\begin{equation}
    \label{eq:zero_bh_embb_q0}
    R(\rho) = \sqrt{\frac{\pi L^2 b}{2}} m_e
    \frac{e^{-\frac{b L^2}{\rho }}}{\sqrt{\rho}}
    \left[1  +\mathcal{O} (\rho^2) \right],
\end{equation}
which agree with (3.6a) and (3.11) in \cite{BitaghsirFadafan:2020lkh}, respectively.

\subsection{Free energy and phase diagram}
\begin{figure}[ht]
\begin{center}
\subfigure[$q/b=1$]{\includegraphics[width=0.4\textwidth]{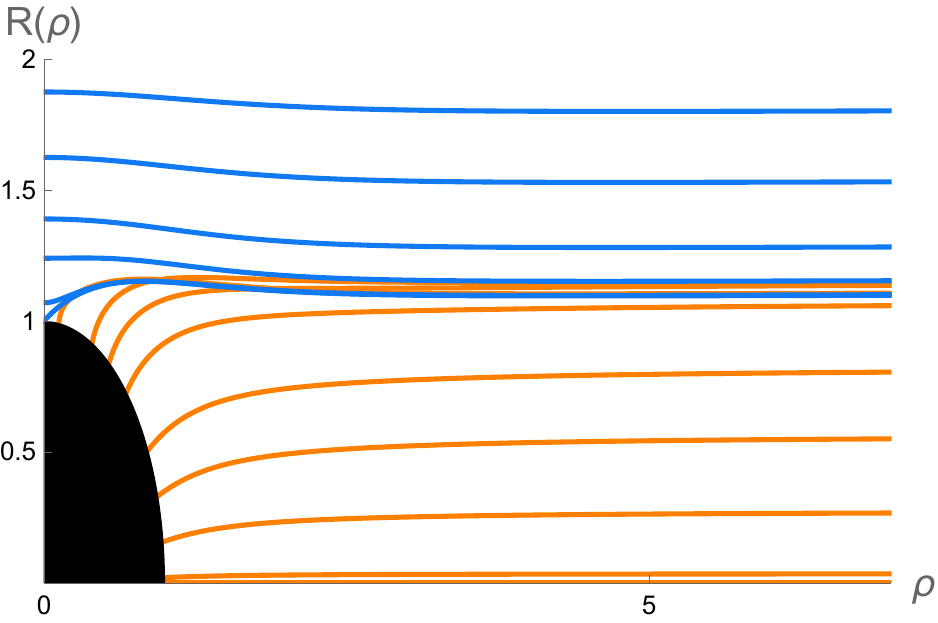}}
\hskip.5cm
\subfigure[$q/b=7$]{\includegraphics[width=0.4\textwidth]{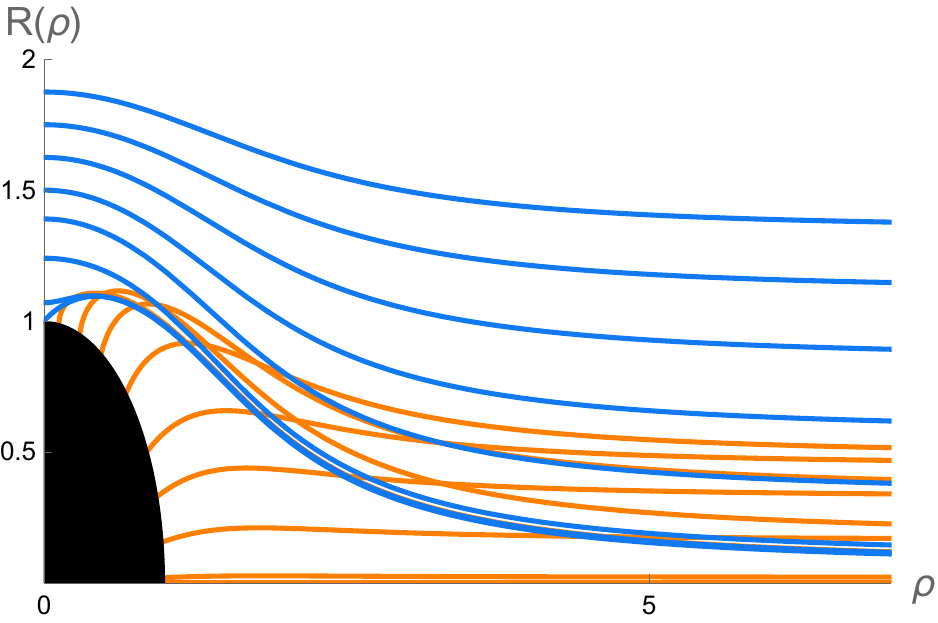}}
\caption{$m_e$ dependence of D7 brane embeddings for (a) $q/b=1$, (b) $q/b=7$. The blue lines denote the Minkowski embeddings and the orange line to black hole embeddings, respectively. Here, we set $\xi_h=1$, which is shown as a black disc.}
\label{fig:D7Embs2}
\end{center}
\end{figure}

In the previous section, we found that the black hole embedding is the only solution for large values of $b$,  while the Minkowski embedding is favored for large values of $q$. 
For generic values of $b$ and $q$, however, there is a competition between the attractive effect of $b$ and the repulsive effect of $q$. Figure \ref{fig:D7Embs2} shows the $m_e$ dependence of the D7 brane embeddings for $q/b=1$ (a) and $q/b=5$ (b).

In the figure, the large value of $m_e$ (or low temperature) region is dominated by the Minkowski embeddings, while the small value of $m_e$ (or high temperature) region is dominated by the black hole embeddings for generic values of $b$ and $q$. In the intermediate region of $m_e$, both types of embeddings coexist. The physically preferred embedding is determined by comparing their free energies.

The Helmholtz free energy $F$ is defined as a negative of the on-shell D7-brane action \eqref{eq:D7_action_potential}
in Euclidean signature. Since the integral diverges as $\rho^4$, it must be regularized by subtracting the contribution from a reference (trivial) solution. 
The resulting free energy density is
\begin{equation}
\label{eq:free_energy}
F= \tau_7 \left[ \int^{\rho_c} d\rho ~ V(\rho,R) \sqrt{1+R'^2}
-\frac{1}{4} \rho_c^4 \right],
\end{equation}
where $\rho_c$ is a large-$r$ cutoff. 

\begin{figure}[ht]
\begin{center}
\subfigure[$q/b=1$]{\includegraphics[width=0.4\textwidth]{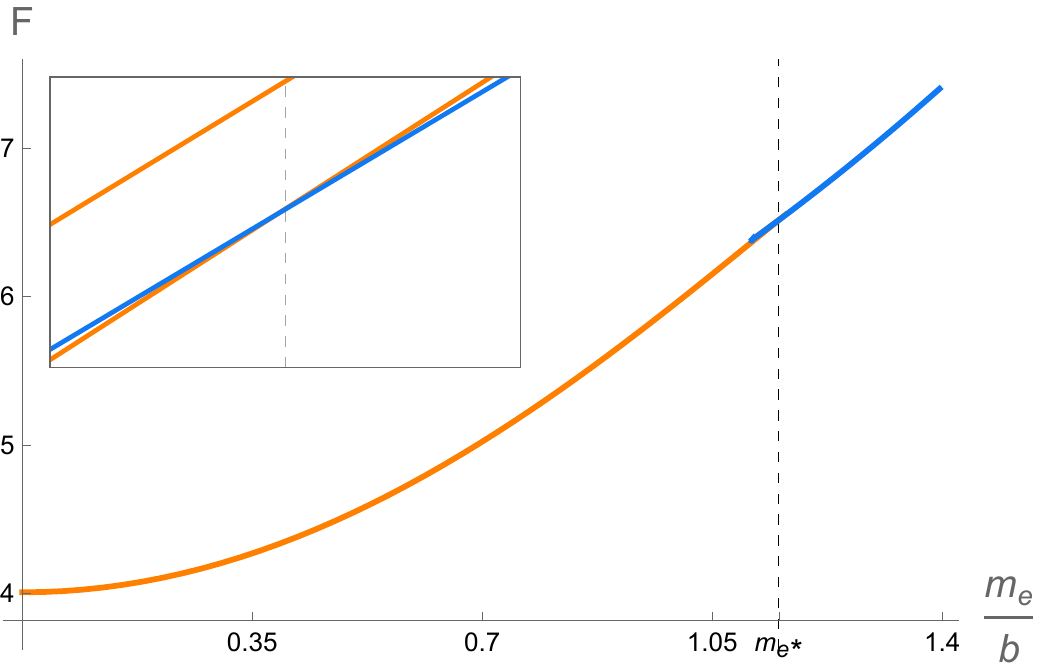}}
\hskip.5cm
\subfigure[$q/b=1$]{\includegraphics[width=0.4\textwidth]{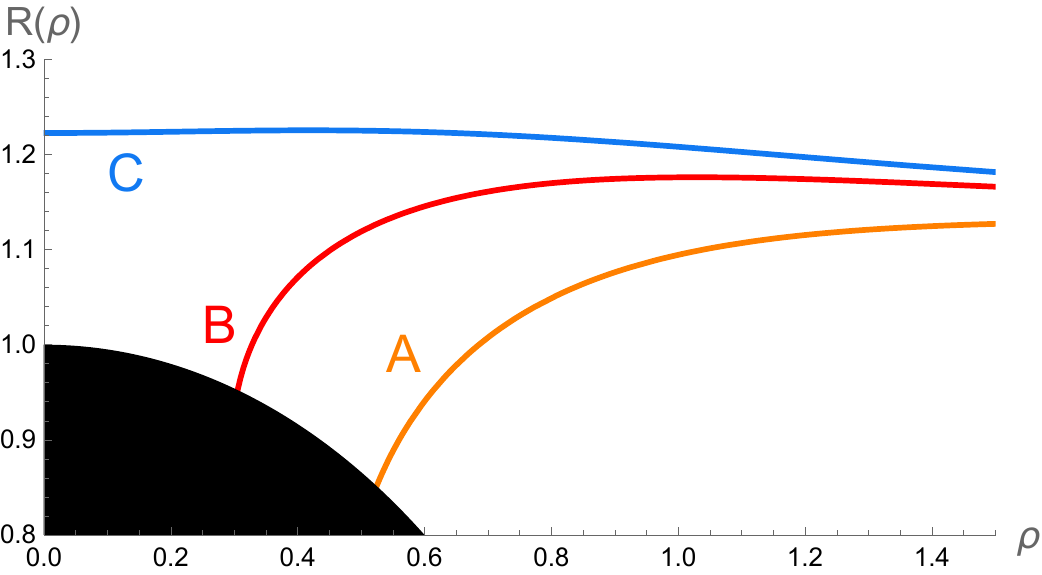}}\\
\subfigure[$q/b=7$]{\includegraphics[width=0.4\textwidth]{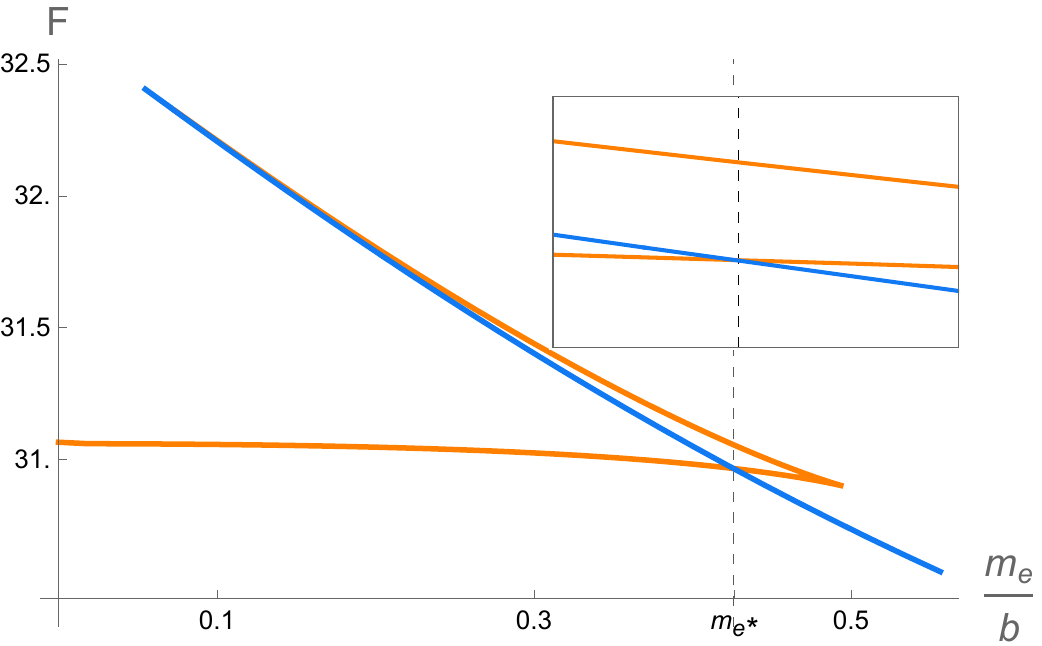}}
\hskip.5cm
\subfigure[$q/b=7$]{\includegraphics[width=0.4\textwidth]{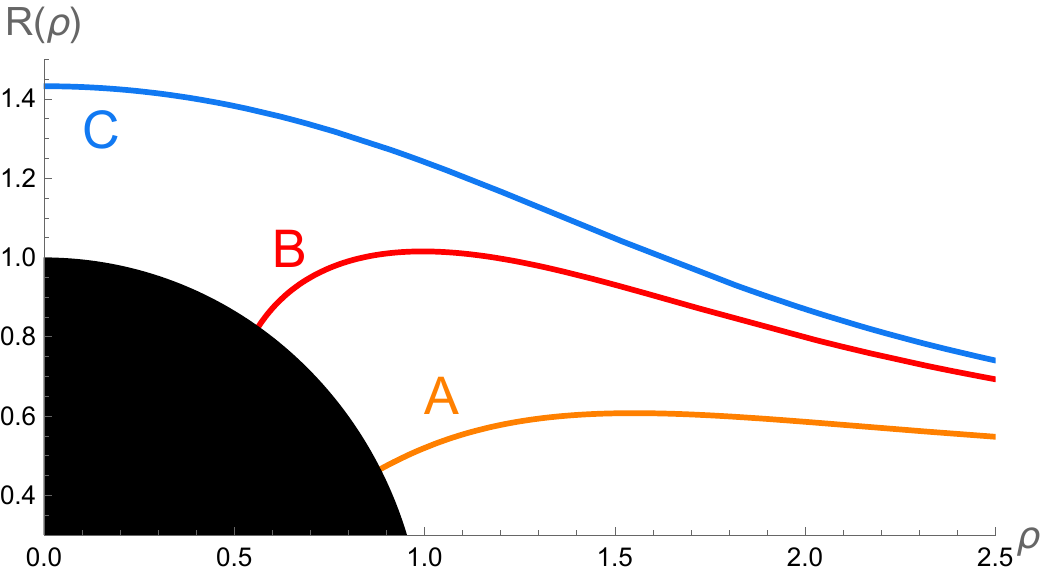}}
\caption{$m_e$ dependence of the free energy density $f$ for (a) $q/b=1$, (c) $q/b=7$. 
Three embeddings for the critical $m_e$ (vertical dashed lines)
are plotted for (b) $q/b=1$, (d) $q/b=7$, respectively.
%The phase transition occurs from the black hole ($A$) to the Minkowski ($C$).
}
\label{fig:freeE}
\end{center}
\end{figure}

Figure \ref{fig:freeE} (a) and (c) show the $m_e$ dependence of the free energy corresponding to the embeddings in Figure \ref{fig:D7Embs2}. As shown in the figure, a first-order phase transition occurs at a critical value of $m_e$. All parameters are scaled by the Weyl parameter $b$. Near the transition point, three types of embeddings exist: one Minkowski embedding and two black hole embeddings, as illustrated in Figure \ref{fig:freeE} (b) and (d).

The free energy indicates that the system is in the black hole embedding phase for $m_e \ll 1$. As $m_e$ increases, a first-order phase transition occurs from the black hole embedding to the Minkowski embedding at a critical value $m_{e}^*/b$. The embedding $B$ between the black hole embedding $A$ and the Minkowski embedding $C$ in Figure \ref{fig:freeE} (b), (d) always has a larger free energy than $A$ or $C$, and so it can never be realized as a physical embedding.

\begin{figure}[ht!]
\begin{center}
\subfigure[]{\includegraphics[width=0.55\textwidth]{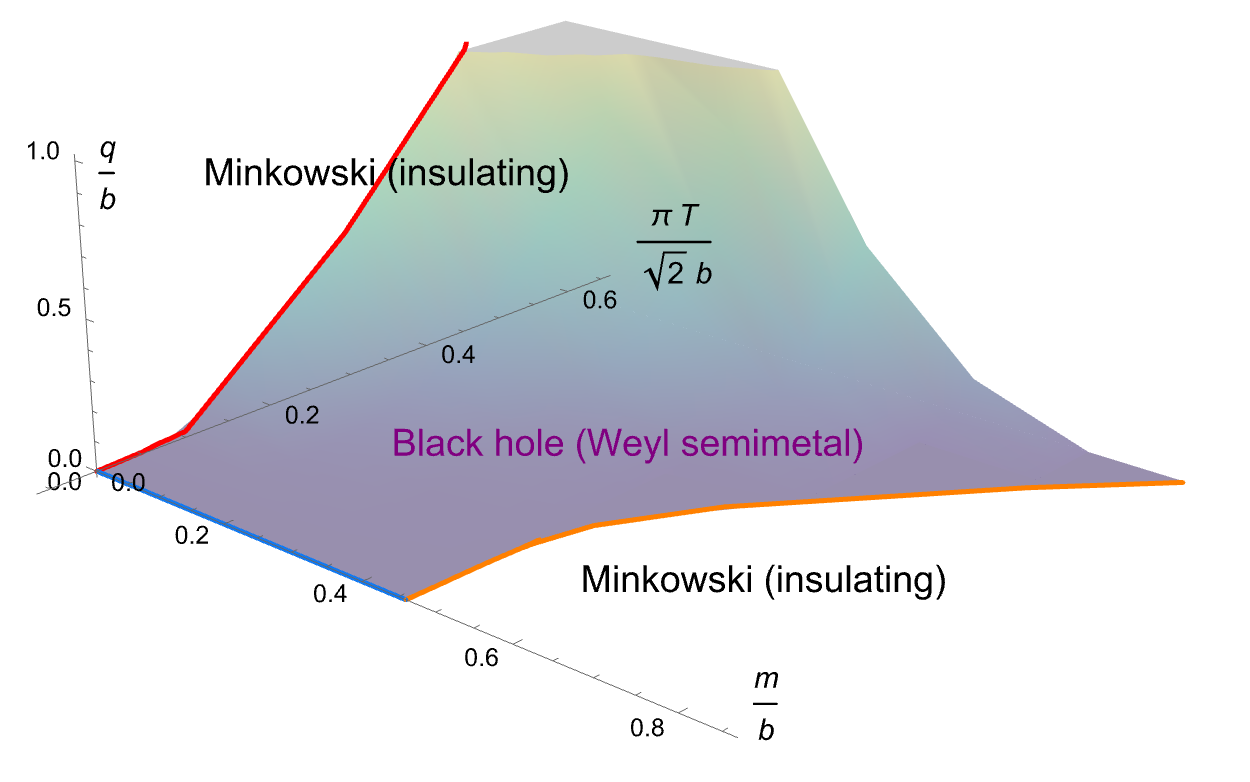}}
\hskip.1cm
\subfigure[]{\includegraphics[width=0.35\textwidth]{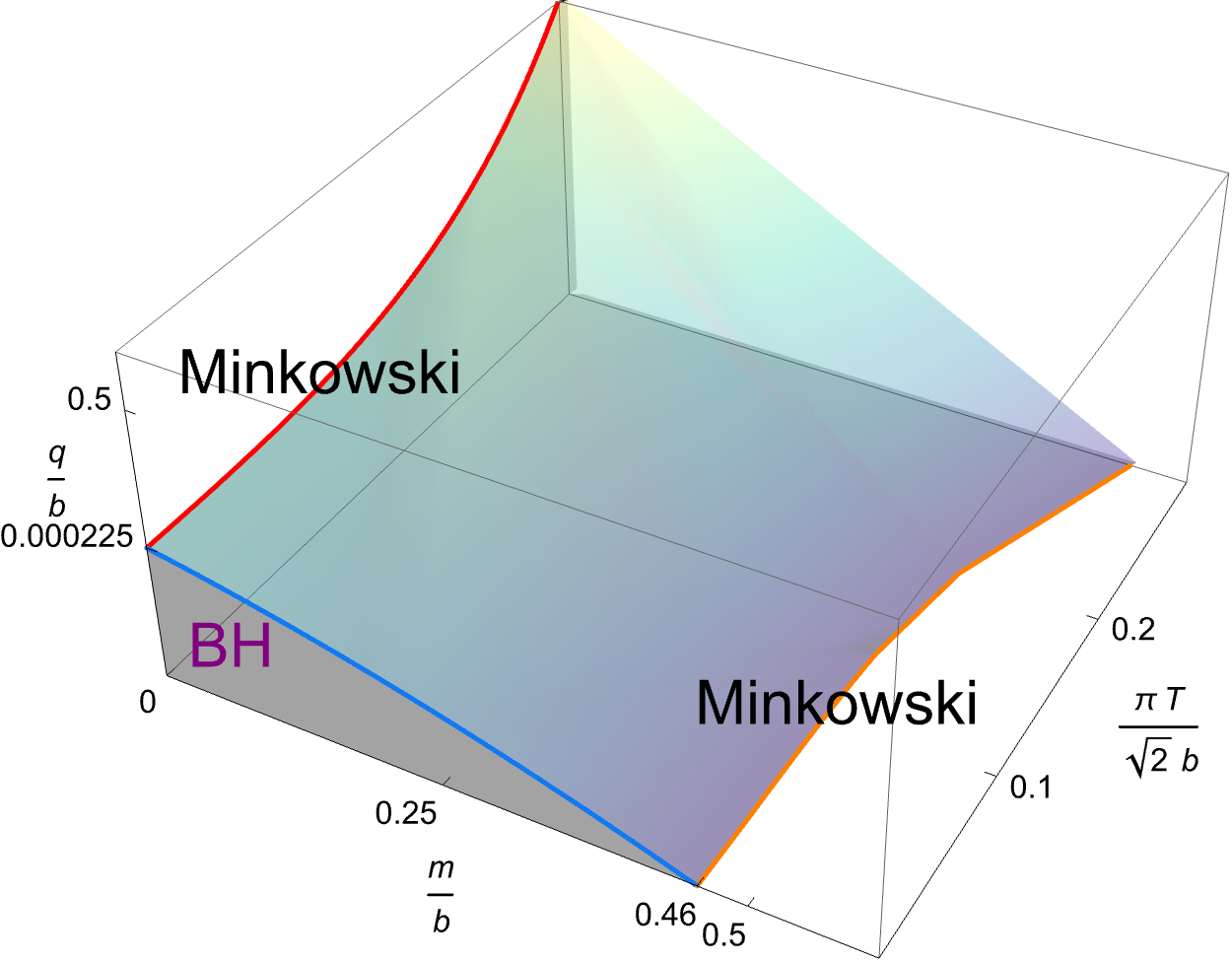}}
\subfigure{\includegraphics[width=0.07\textwidth]{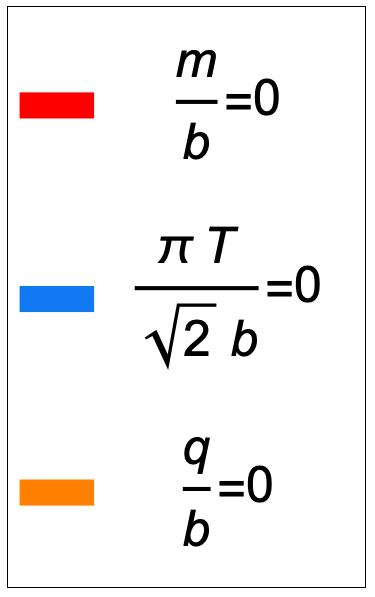}}
\caption{phase diagram surface in $\left( m, T, q\right)$
for a nonzero $b$. The right figure is a detailed phase diagram near the origin of the left phase diagram.}
\label{fig:phase}
\end{center}
\end{figure}

The temperature of the boundary system is defined by
\begin{equation}
\label{eq:temperature}
T= \frac{\sqrt{2}}{\pi L^2} \xi_H.
\end{equation}
The phase diagram in terms of the temperature, electron mass, and the D-instanton density is shown in Figure \ref{fig:phase}. All quantities are scaled by the Weyl parameter $b$. In the figure, the region with small $m_e/b$ and the small $q/b$ is dominated by the black hole embedding, while the Minkowski embedding phase exists outside of the black hole embedding region. 

The boundary state corresponding to the black hole embedding can be identified as a metallic phase via the nonlinear conductivity, which will be discussed in the next section. In the Figure \ref{fig:phase} (a), the phase structure in the $q/b =0$ plane agrees with the result of \cite{BitaghsirFadafan:2020lkh}. The detailed structure near the origin ($T/b \ll 1$, $m/b \ll 1$ and $q/b\ll 1$) is drawn in Figure \ref{fig:phase} (b). As shown in the figure, even though we start with a Weyl semimetal phase at $q/b=0$, there is a phase transition to the insulating phase for a large value of $q/b$. This phenomenon indicates that the instanton density generates a bulk gap, driving the system from a Weyl semimetal to an insulator.

The last discussion in this section concerns the embeddings in the $m_e=0$ case. In the absence of the instanton density $q$, the only solution for $m_e=0$ is $R(\rho)=0$ for the entire region $\rho>\rho_H$. Therefore, the D7-brane embedding is of the black hole type. This is also shown in Figure \ref{fig:phase}. For this black hole embedding, the chiral condensation defined in (\ref{eq:asym_D7}) vanishes.

However, as the instanton density $q$ increases, the D7-brane experiences a repulsive force near the black hole horizon, as shown in Figure \ref{fig:D7Embs} (d), and eventually transitions to a Minkowski embedding. From the analysis of the free energy, we find that there is a first-order phase transition from the black hole embedding to the Minkowski embedding, similar to the finite-$m_e$ case.

\begin{figure}[ht]
\begin{center}
\subfigure[]{\includegraphics[width=0.4\textwidth]{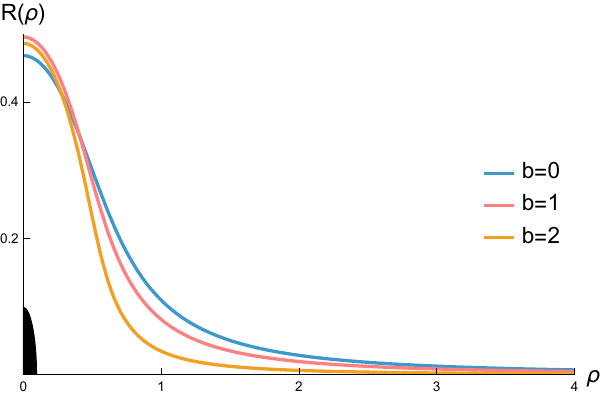}}
\hskip.5cm
\subfigure[]{\includegraphics[width=0.5\textwidth]{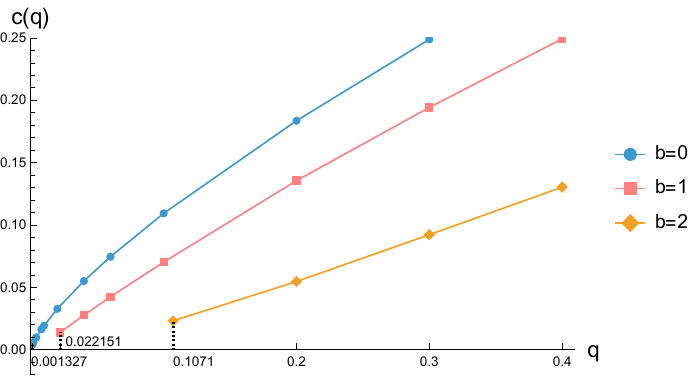}}
\caption{(a) D7-brane embeddings for $m_e=0$ with different value of $b$ for fixed value of $q\sim 0.1$. (b) $q$ dependence of the chiral condensation for $m_e =0$ for different value of $b$. Both case, we set $\xi_h =0.1 .$}
\label{fig:D7Embs_m0}
\end{center}
\end{figure}

Figure \ref{fig:D7Embs_m0} (a) shows examples of Minkowski embeddings for $m_e=0$. In this case, the chiral condensation is clearly nonzero due to the finite slope of the D7-brane embedding at infinity. The dependence of the chiral condensation on the instanton number density is shown in Figure \ref{fig:D7Embs_m0} (b). For small values of $q$, the black hole embedding with $c=0$ is the physical solution. However, as the instanton number density increases, the system undergoes a transition to the Minkowski embedding, and the chiral condensation consequently jumps to a finite value. As $q$ increases further, the repulsive effect becomes stronger, leading to an increase in the chiral condensation.

In the figure, one can also notice that the parameter $b$ suppresses the chiral condensation. This is due to the attractive effect of the Weyl parameter $b$ on the probe D7-brane, as discussed above.

\begin{figure}[ht!]
\begin{center}
\includegraphics[width=0.55\textwidth]{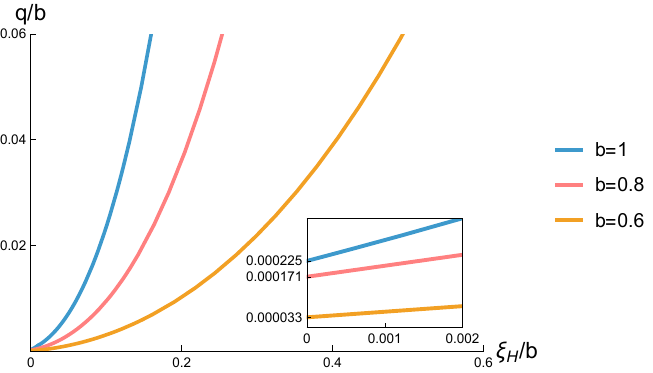}
\caption{$b$ dependence of phase transition temperature at $m_e/b=0$. }
\label{fig:phase_for_b}
\end{center}
\end{figure}

Figure \ref{fig:phase_for_b} shows the dependence of the phase transition on the Weyl parameter $b$ at $m_e/b=0$. The horizontal axis corresponds to $q/b$, and the line $q/b=0$ is always in the black hole(WSM) phase. As $b$ increases,  the transition from the Weyl semimetal to the insulating phase occurs at larger values of  $q/b$ (from the yellow line to the blue line in the figure). This is consistent with our earlier observation: the Weyl parameter $b$ favors the Weyl semimetal phase, while the instanton density $q$ drives the system toward an insulating phase.

\section{Conductivity}
\subsection{Nonlinear electric currents}
Now, we calculate non-linear DC conductivity using the method of \cite{Karch:2007pd,Hashimoto:2014yya}. To this end, we introduce gauge field fluctuations corresponding to an external electric field in the $x$ direction. From the asymptotic behavior of the gauge field, the boundary currents $\left<J_x\right>$ and $\left<J_y\right>$ can be obtained. The regularity condition at the black hole horizon provides a relation between the external electric field and the boundary currents. By extracting the coefficients of the linear terms in the external electric field, we obtain the nonlinear longitudinal and Hall conductivities.

In this section, we study the dependence of the longitudinal and transverse currents on the external electric field. The DC conductivities are obtained in the next section by taking the zero-field limit.

We consider the following ansatz for the $U(1)$ gauge field on the D7-brane:
\begin{equation}
\label{eq:DC_setting}
A_x (t,\rho) =E t+A_x(\rho) ,
\quad
A_y (\rho) ,
\quad
\phi(z,\rho) =bz +\phi(\rho).
\end{equation}
Note that
we introduce the electric field $E$ in the $x$ direction,
and
$A_i(\rho)$
is allowed for a nonzero current
$\langle J_i \rangle$ for $i=x,y$.
%We use rotational symmetry in the $xy$-plane to set $E>0$.
From
the ansatz \eqref{eq:DC_setting},
the metric and the pullback of $C_4$ to the worldvolume are given by
\begin{align}
\label{eq:metric_WSM_DC}
ds^2_{D7} = \! e^{\Phi/2} \!
\Big[
\frac{r^2}{L^2}
&
\left(
-\frac{\omega^2_-}{\omega^2_+} dt^2
+\left(dx^2 +dy^2 \right) \right)
+\left(\frac{r^2}{L^2} +\frac{L^2 b^2 R^2}{\xi^2} \right) dz^2 \cr
&+\frac{L^2}{\xi^2}
\left(
\left(1+R'^2 +R^2 \phi^{'2} \right)d\rho^2
+\rho^2 d\Omega^2_3 
\right)
\Big],
\end{align}

\begin{equation}
\label{eq:C4_WSM_DC}
P[C_4 ] =\frac{\xi^4}{L^4} \omega_+^2 dt \wedge dx\wedge dy\wedge dz
-\frac{L^4 \rho^4}{\xi^4}
\left(b~dz+\phi' d\rho\right) \wedge d\Omega_3 ,
\end{equation}
and then,
the Wess–Zumino term becomes
\begin{equation}
S_{WZ} =\frac{1}{2}
N_f \mu_7 \int P[C_4 ]
\wedge F \wedge F
\sim \int d\rho \frac{L^4 \rho^4}{\xi^4} b E A_y' ,
\end{equation}
where $2\pi\alpha'$ has been absorbed into $F$.

Then, the DBI action \eqref{eq:D7_action} becomes ($\mathcal{N}=\tau_7 ~{\rm vol}(\mathbb{R}^{(1,3)})$)
\begin{equation}
\label{eq:action_DC}
S_{D7} =  -\mathcal{N}
\int d\rho
\left(
\sqrt{w_1 \left[ \omega_+ \left(1+R'^2 \right) +A_y^{'2} \right] + w_2 \phi^{'2}+w_3 A_y^{'2}}
-w_4 Ay^{'}
\right),
\end{equation}
where
\begin{align}\label{eq:w-functions}
w_1 (\rho) = \rho^6  \left(\frac{L^4 b^2 R^2}{\xi^4} +\omega_+ \right)
\left(e^\Phi \omega_-^2 -\frac{L^4 E^2}{\xi^4}\right), &
~~~ w_2 (\rho) = \rho^6 
e^\Phi R^2 \omega_+^2\left(e^\Phi \omega_-^2 -\frac{L^4 E^2}{\xi^4}\right),\cr
 w_3 (\rho) = \rho^6 
e^\Phi \omega_-^2 \left(\frac{L^4 b^2 R^2}{\xi^4} +\omega_+ \right),
\, &  ~~~
w_4 (\rho) = \frac{L^4 E b \rho^4}{\xi^4} . 
\end{align}
We define the canonical momenta 
$P_\phi \equiv \delta S_{D7} /\delta \phi'$,$P_x \equiv \delta S_{D7} /\delta A_x'$, and $P_y \equiv \delta S_{D7} /\delta A_y'$ to $\phi$, $A_x$, and $A_y$, respectively.
Then the Euler-Lagrange equations are given as $\partial_\rho P_i =0$ ($i=\phi,x,y$),
which implies that
the canonical momenta can be written as
some constant such as
$P_\phi =\mathcal{N} p_\phi$,
$P_x =\mathcal{N} j_x$,
$P_y =\mathcal{N} j_y$.

By using the Legendre transformation
with respect to $\phi$, $A_x$, and $A_y$,
we finally get
\begin{align}
\notag
\tilde{S}_{D7} &
=
S_{D7} -\mathcal{N} \int d\rho
\left(\phi' P_\phi + A_x' P_x
+A_y' P_y \right) \\\notag
& = -\mathcal{N}
\int d\rho \sqrt{\omega_+} 
\sqrt{1+R^{'2}} \sqrt{\alpha(\rho) \beta(\rho) -\gamma(\rho) },
\end{align}
where
\begin{align}
\label{eq:cond_func}
&
\alpha(\rho) = e^\Phi \omega_-^2 -\frac{L^4 E^2}{\xi^4},
\quad
\beta(\rho) =\rho^6 \left(\frac{L^4 b^2 R^2}{\xi^4} +\omega_+ \right)
-  \frac{j_x^2}{e^\Phi \omega_-^2},
\\\notag
&
\gamma(\rho)
=
\left(\frac{L^4 b^2 R^2}{\xi^4} +\omega_+ \right)
\frac{p_\phi^2}{\omega_+^2 e^\Phi R^2} 
+e^\Phi \left(\frac{L^4 b E \rho^4}{\xi^4} -j_y \right)^2.
\end{align}
Note that $\gamma(\rho)$ is nonnegative for all $\rho$,
whereas
the signs of $\alpha(\rho)$ and $\beta (\rho)$ are flipped at some $\rho_H < \rho <\infty$.
In order that $\alpha(\rho) \beta(\rho) -\gamma(\rho) \geq 0$
in the range of $\rho \geq\rho_H$,
the roots of $\alpha(\rho)$, $\beta(\rho)$, and $\gamma(\rho)$ should be same each other.
We denote the root as $\rho_*$
which corresponds to the so-called worldvolume horizon.

Denoting the values of $R$, $\xi$,  $\omega_\pm$ and $\Phi$ at $\rho_*$
as
$R_*$, $\xi_*$, $\omega_{\pm *}$, and $\Phi_*$, respectively,
we can write $\alpha(\rho_*) =\beta(\rho_*) =0$ in \eqref{eq:cond_func} as
\begin{equation}
\label{eq:alpha_vanish}
e^{\Phi_*} \omega_{-*}^2 -\frac{L^4 E^2}{\xi^4_*} =0,
\end{equation}
\begin{equation}
\label{eq:beta_vanish}
\rho^6_* \left(\frac{L^4 b^2 R^2_*}{\xi^4_*} +\omega_{+*} \right)
-  \frac{j_x^2}{e^{\Phi_*} \omega_{-*}^2} =0.
\end{equation}
From the equations, we get the longitudinal quantity $j_x$ as 
\begin{equation}
\label{eq:cond_eq_x}
j_x = -
\frac{L^2 \rho^3_* E}{\xi^2_*}
 \sqrt{\frac{L^4 b^2 R^2_*}{\xi^4_*} +\omega_{+*} } .
\end{equation}
In order that $\gamma(\rho_* )=0$,
each term in $\gamma(\rho_* )$ should vanish independently.
\begin{equation}
\label{eq:cond_eq_y}
p_\phi =0,
\quad
j_y = 
\frac{L^4 b E \rho^4_*}{\xi^4_*}
\end{equation}

Currents would be a conjugate momentum to the source of the external gauge field, $i.e.$, the electric field $E$. Therefore, it is natural to define electric currents as
\begin{equation}
\label{eq:J_def}
\langle J_x \rangle = -(2\pi\alpha') \mathcal{N} j_x ,
\qquad
\langle J_y \rangle = -(2\pi\alpha') \mathcal{N} j_y.
\end{equation}
From \eqref{eq:cond_eq_x} and \eqref{eq:cond_eq_y}, we get electric currents in terms of external electric field $E$ as follows;
\begin{equation}
\label{eq:J_func}
\langle J_x \rangle = (2\pi\alpha') \mathcal{N}
\frac{L^2 \rho^3_* E}{\xi^2_*}
 \sqrt{\frac{L^4 b^2 R^2_*}{\xi^4_*} +\omega_{+*} } ,
 \quad
 \langle J_y \rangle =
 -(2\pi\alpha') \mathcal{N}
 \frac{L^4 b E \rho^4_*}{\xi^4_*} .
\end{equation}

Note that $\xi_*$ is determined by \eqref{eq:alpha_vanish} for a given value of $E$, and it is nothing but the world volume horizon position on the probe D7-brane with $\xi_*^2 = \rho_*^2+R_*^2$. One can easily check that $\xi_* \rightarrow \xi_H$ when the external electric field vanishes and the value of $\xi_*$ is proportional to the external electric field $E$. 

In the case of the black hole embedding, the D7-brane always touches the black hole horizon, and hence the worldvolume horizon exists for any value of the external electric field. On the other hand, the Minkowski embedding does not touch the black hole horizon, and there is a finite separation between the D7-brane and the horizon. If the external electric field is not sufficiently large, the value of $\xi_*$ can be smaller than this minimal distance. In this case, (\ref{eq:J_def}) are no longer valid, and both electric currents $\langle J_x \rangle$ and $\langle J_y\rangle$ vanish. However, if the external electric field increases enough such that $\xi_*$ exceeds the minimal distance to the D7-brane, a worldvolume horizon forms on the D7-brane embedding, and electric currents are generated according to (\ref{eq:J_def}).

\begin{figure}[ht]
\begin{center}
\subfigure[$q=1$]{\includegraphics[width=0.49\textwidth]{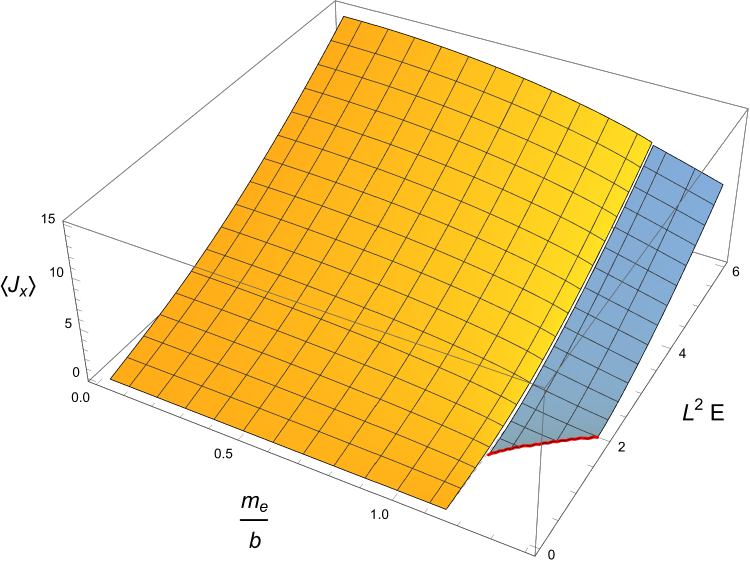}}
\hskip.1cm
\subfigure[$q=7$]{\includegraphics[width=0.49\textwidth]{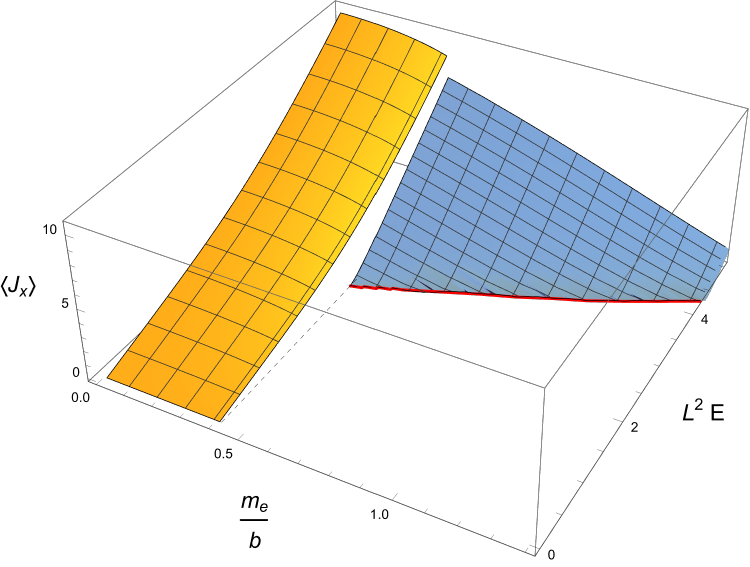}}
\caption{Nonlinear electric currents $\langle J_x \rangle$ in \eqref{eq:J_func} for $\xi_H =1$ and $b=1$.  We set $T/b = 0.45$. Yellow surfaces represent the electric current of the black hole embedding, and the blue surfaces are for the Minkowski embedding.
}
\label{fig:nonlin}
\end{center}
\end{figure}

The numerical results for the dependence of the longitudinal current on the external electric field and the electron mass are shown in Figure \ref{fig:nonlin}. In the figure, the yellow surface corresponds to the current in the black hole embedding, while the blue surface corresponds to that in the Minkowski embedding for different values of the instanton density $q$. 

In the black hole embedding, the current is immediately generated by the external electric field, which is similar to electron-hole pair creation near the Fermi surface. Each electron and hole moves in opposite directions under the applied electric field, generating a finite electric current. The yellow surfaces in Figure \ref{fig:nonlin} represent this behavior, indicating that the boundary system in the black hole embedding corresponds to a metallic phase.

On the other hand, if there is a gap in the electron energy spectrum, electron-hole pairs cannot be excited when the excitation energy is smaller than the gap scale. Such a state can be regarded as a band insulator in condensed matter physics. However, when the external electric field exceeds the gap energy, electron-hole pair creation becomes possible. The blue surfaces in Figure \ref{fig:nonlin} correspond to this situation. In the Minkowski embedding, the electric current is not generated below a certain value of the external electric field, but once the electric field exceeds a critical value, the current is generated along the blue surface.

\begin{figure}[ht]
\begin{center}
\includegraphics[width=0.54\textwidth]{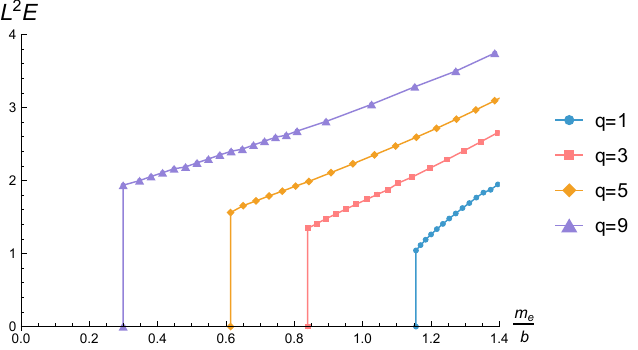}
\caption{Critical values of $E$ for $\xi_H =1$ and $b=1$. We set $T/b = 0.45$.
}
\label{fig:critical_E}
\end{center}
\end{figure}

This critical electric field for different instanton numbers is shown as the red line in Figure \ref{fig:critical_E}, where the current starts to be generated in the Minkowski embedding.

Here, we have focused on the dependence of the longitudinal electric current on the external electric field. The transverse electric current also exhibits a nontrivial dependence on the external electric field through \eqref{eq:J_func}. We find that its overall behavior is qualitatively similar to that of the longitudinal current.

\subsection{DC conductivity}
One of the key electrical properties of a material at the linear response level is the DC conductivity. Whether the longitudinal DC conductivity is zero or not determines whether the material is in a metallic or insulating phase. A nonzero transverse DC conductivity indicates spontaneous magnetization or the anomalous Hall effect. We can compute the DC conductivities from the electric currents by taking the zero electric field limit as
\begin{equation}
\label{eq:DC_cond_def}
\sigma_{xx} = \lim_{E\rightarrow 0} (2\pi\alpha') \langle J_x \rangle /E ,
\qquad
\sigma_{xy} =-\sigma_{yx} = - \lim_{E\rightarrow 0} (2\pi\alpha') \langle J_y \rangle /E .
\end{equation}
From \eqref{eq:J_func}
and the temperature definition \eqref{eq:temperature}, we obtain the longitudinal and Hall DC conductivities as functions of the horizon values of the D7-brane embedding:
\begin{align}
\label{eq:DC_cond_func}
& \sigma_{xx} = 
(2\pi\alpha')^2 \mathcal{N}
\frac{L^2 \rho^3_H}{\xi^2_H}
 \sqrt{\frac{L^4 b^2 R^2_H}{\xi^4_H} + 2 }
= \frac{(4\alpha')^2 \mathcal{N} \rho^3_H}{\pi^2 L^4 T^4}
 \sqrt{\frac{1}{2} \pi^4 L^4 T^4 +b^2 R_H^2}
 ,\\ \label{eq:DC_cond_func2}
& \sigma_{xy} =
 (2\pi\alpha')^2 \mathcal{N} \frac{L^4 b \rho^4_H}{\xi^4_H} 
 = \frac{(4\alpha')^2 \mathcal{N} b \rho^4_H}{\pi^2 L^4 T^4}.
\end{align}
Note that in the zero electric field limit, the worldvolume horizon always coincides with the black hole horizon. Therefore, the DC conductivities for the Minkowski embedding vanish, and the expressions in \eqref{eq:DC_cond_func} and \eqref{eq:DC_cond_func2} are valid only for the black hole embedding.

When $m=0$, the trivial solution ($\rho_h = \xi_H$ and $R_H=0$) is always thermodynamically preferred. In this case, the conductivities reduce to
\begin{equation}
\label{eq:DC_cond_m0}
\sigma_{xx}|_{m=0} =  (2\alpha')^2 \pi^3  L^4 \mathcal{N} ~ T,
\qquad
\sigma_{xy}|_{m=0} =  (2\alpha')^2 \pi^2  L^4 \mathcal{N}~ b.
\end{equation}

\begin{figure}[ht!]
\begin{center}
\subfigure[$\sigma_{xx}$]{\includegraphics[width=0.4\textwidth]{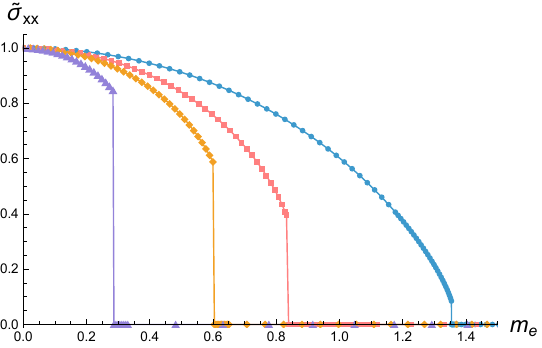}}
\hskip.1cm
\subfigure[$\sigma_{xy}$]{\includegraphics[width=0.48\textwidth]{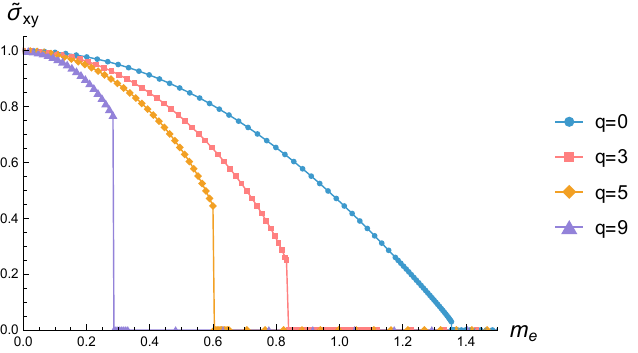}}
\subfigure[$\sigma_{xx}$]{\includegraphics[width=0.4\textwidth]{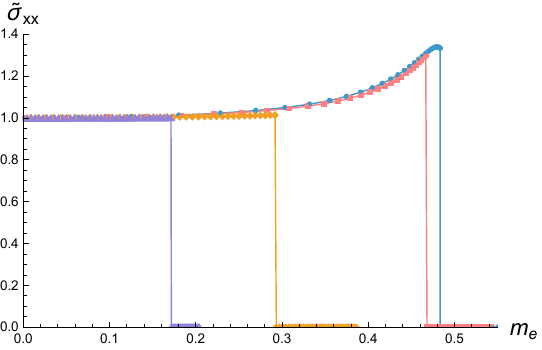}}
\hskip.1cm
\subfigure[$\sigma_{xy}$]{\includegraphics[width=0.48\textwidth]{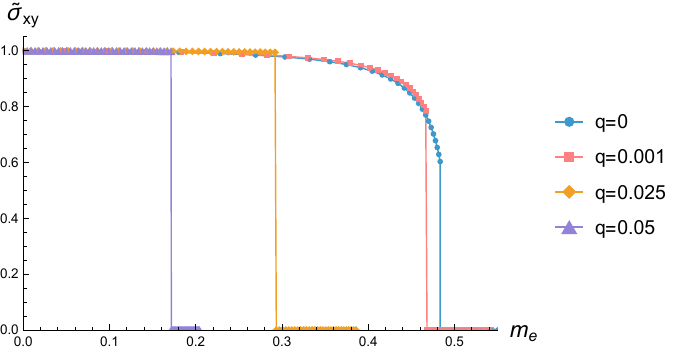}}
\caption{(a) and (b): Normalized conductivities for $\xi_H =1$ and $b=1$ with $T/b \approx 0.45$. (c) and (d):Normalized conductivities for $\xi_H =0.2$ and $b=1$ with $T/b \approx 0.09$.
}
\label{fig:cond}
\end{center}
\end{figure}

For simplicity, we consider the conductivities normalized by \eqref{eq:DC_cond_m0} such as
\begin{equation}
\label{eq:DC_cond_normal}
\tilde{\sigma}_{xx} \equiv \frac{\sigma_{xx}}{\sigma_{xx}|_{m=0}}
\sim \frac{\sigma_{xx}}{T},
\qquad
\tilde{\sigma}_{xy} \equiv \frac{\sigma_{xy}}{\sigma_{xy}|_{m=0}}
\sim \frac{\sigma_{xy}}{b}.
\end{equation}
These normalized conductivities in terms of $m_e$ are plotted in Figure~\ref{fig:cond}.

Figure \ref{fig:cond} (a) and (b) show the normalized longitudinal and Hall conductivities, respectively. As discussed earlier, finite DC conductivities appear in the black hole embedding for small electron mass at finite temperature (here we set $\xi_H=1)$. As the electron mass $m_e$ increases, both $\sigma_{xx}$ and $\sigma_{xy}$ decrease monotonically. When $m_e$ becomes sufficiently large, a phase transition to the Minkowski embedding occurs, and the system enters an insulating phase. This first-order phase transition is reflected as a sudden drop in the conductivities.

Figure \ref{fig:cond} (c) and (d) show the conductivities at low temperature (we set $\xi_H=0.2$). In this regime, a small peak appears in the longitudinal conductivity just below the phase transition point, as already observed in \cite{BitaghsirFadafan:2020lkh}. However, this peak is suppressed when the instanton density is turned on, and the instanton reduces both the longitudinal and Hall conductivities in the Weyl semimetal phase.

\begin{figure}[ht!]
\begin{center}
\subfigure[Longitudinal conductivity $\sigma_{xx}$]{\includegraphics[width=0.45\textwidth]{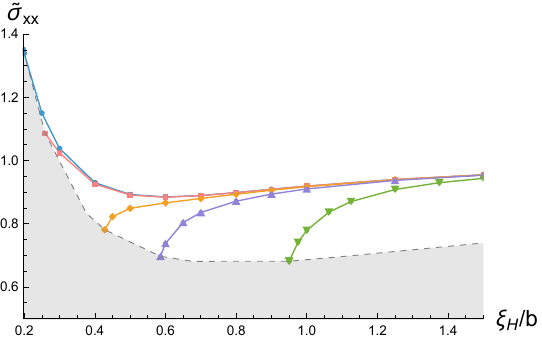}}
\hskip.1cm
\subfigure[Hall conductivity $\sigma_{xy}$]{\includegraphics[width=0.53\textwidth]{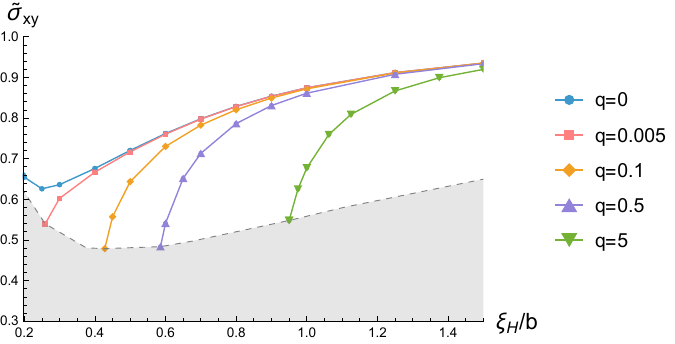}}
\caption{Normalized conductivities for $m_e=0.48$ and $b=1$.
The shaded region indicates the Minkowski phase.
}
\label{fig:cond-temp}
\end{center}
\end{figure}

Finally, we analyze the temperature dependence of the conductivities. For a given electron mass and Weyl parameter, the temperature dependence of the longitudinal and Hall conductivities is shown in Figure~\ref{fig:cond-temp}. In the regime where the black hole horizon scale is comparable to the electron mass, the D7-brane embedding has a nontrivial profile, and the conductivities exhibit nontrivial temperature dependence.

When the horizon scale becomes much larger than the electron mass, the temperature dominates, and the conductivities approach the massless limit given in \eqref{eq:DC_cond_m0}. Accordingly, the normalized conductivities approach constant values. This behavior is also evident in Figure~\ref{fig:cond-temp}, where all curves tend to converge at high temperature.

\section{Conclusion and Discussion}

\begin{figure}[ht!]
\begin{center}
\includegraphics[width=0.8\textwidth]{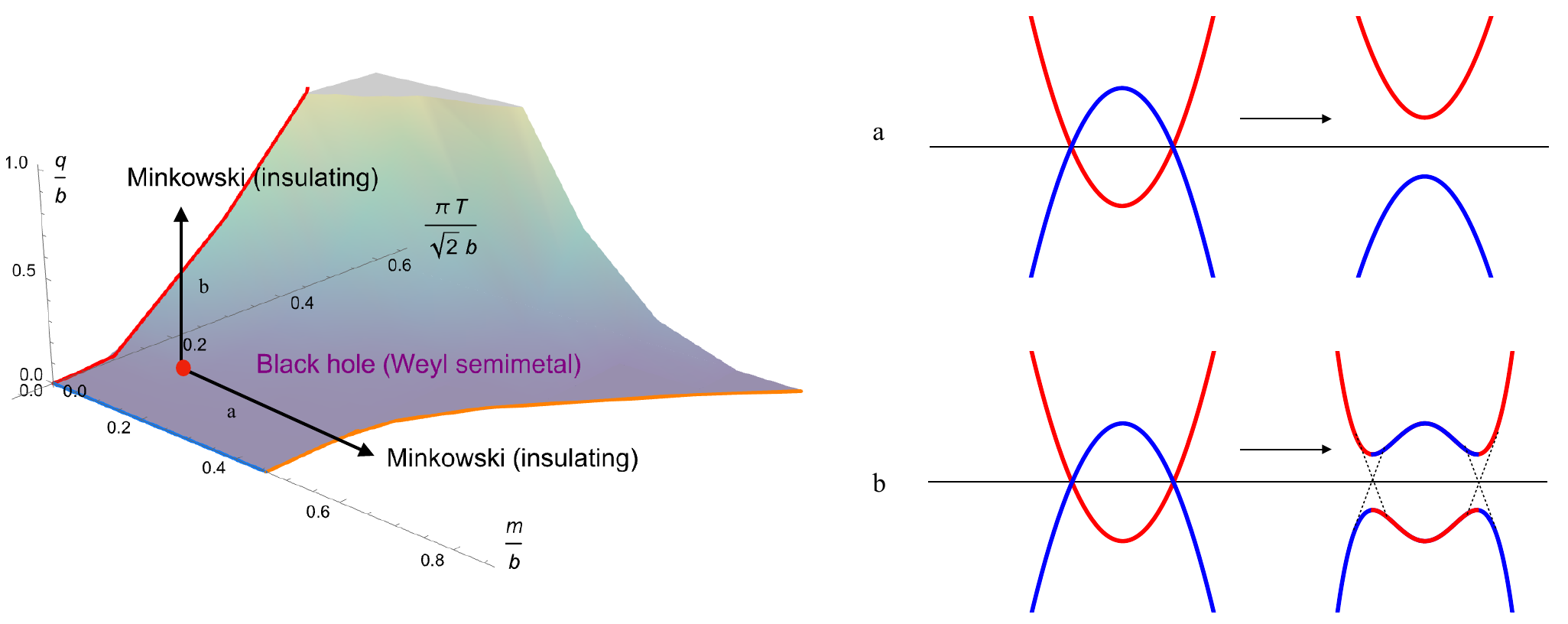}
\caption{Two possible types of the WSM to the insulating phase transition. }
\label{fig:top_transition}
\end{center}
\end{figure}

In this work, we have investigated the effects of D-instantons on a holographic Weyl semimetal in a top-down approach. We found that the Weyl parameter tends to pull the probe D7-brane toward the black hole horizon, while the instanton density generates an effective repulsive force that pushes the D7-brane away from the horizon. By analyzing the free energy of the D7-brane embeddings, we constructed the phase diagram in terms of the fermion mass, instanton density, and temperature, measured in units of the Weyl parameter.

We showed that the black hole embedding is thermodynamically favored when the fermion mass and instanton density are small compared to the Weyl parameter. In contrast, the Minkowski embedding dominates in the regime of large fermion mass or large instanton density. From the boundary theory perspective, the black hole embedding corresponds to a metallic phase identified with the Weyl semimetal, while the Minkowski embedding corresponds to a gapped phase. Therefore, both a large fermion mass and a large instanton density can induce a gap in the boundary theory.

Although the transition from black hole embedding to Minkowski embedding appears similar at the level of the D7-brane geometry, its interpretation in terms of fermionic excitations can be qualitatively different depending on whether it is driven by the fermion mass or by the instanton density. A schematic illustration of these two possibilities is shown in Figure \ref{fig:top_transition}.

In the absence of instantons, the Weyl semimetal phase arises from a band inversion mechanism described by the fermionic effective action (\ref{eq:action_Dirac}), which reduces to the Weyl fermion Lagrangian (\ref{WeylL}). There are two distinct paths toward a gapped phase. One is driven by increasing $m/b$, denoted as path `a' in Figure \ref{fig:top_transition}. The other is driven by increasing the instanton density $q$, corresponding to path `b'.

The gap formation along path `a' can be understood as the inverse process of band inversion, as described by the dispersion relation (\ref{energyE}). In this case, the Weyl nodes approach each other and annihilate, leading to a trivial insulating phase.

In contrast, along path `$b$', the ratio $m/b$ remains fixed. The parameter $m$ represents the distance between the D3- and D7-branes in the asymptotic region and appears as $m_e$ in (\ref{eq:action_Dirac}). This means that the effective mass in the fluctuation Lagrangian (\ref{eq:action_Dirac}) is kept fixed. Therefore, the Weyl band structure is maintained up to the transition point. The gap then opens without the conventional annihilation of Weyl nodes. To preserve the band inversion structure, the conduction and valence bands must remain connected in a nontrivial way, as illustrated in Figure \ref{fig:top_transition}. From a geometric perspective, this process originates from the change in the induced metric on the D7-brane, rather than from a change in the effective mass of the fluctuation fermion. This behavior suggests a qualitatively different mechanism of gap formation from the conventional mass-driven case.

This indicates that the gap induced by the instanton density cannot be fully captured by the conventional Dirac theory (\ref{WeylL}), and may be associated with a nontrivial restructuring of the underlying fermionic spectrum. Instanton effects are known to encode nonperturbative phenomena in condensed matter systems, such as quantum phase transitions and topological defects. Recent studies have also explored instanton-induced effects in topological phases \cite{park2020higherordertopologicalinstantontunneling}.

In our holographic setup, the boundary of $AdS_5$ corresponds to the bulk of a $(3+1)$-dimensional material. Depending on the parameters, the boundary system exhibits a metallic phase identified as a Weyl semimetal, as well as two distinct gapped phases. One corresponds to a trivial insulator induced by a large fermion mass compare to the Weyl parameter `b', while the other, induced by the instanton density,  represents a qualitatively different gapped phase.

A more precise characterization of the instanton-induced phase, including its possible topological nature, requires further investigation, such as the analysis of boundary states or symmetry properties. We leave this for future work.

\section*{Acknowledgment}
We thank Sang-Jin Sin and Keunyoung Kim for useful discussions.
This work is supported by the Basic Science Research Program through the National Research Foundation of Korea(NRF)
grant No. NRF-2022R1A2C1010756.
YS was partially supported by NRF grant funded by the Korea government(MSIT) No. RS-2026-25588089.
HE was partially supported by Basic Science Research Program through the NRF funded by the Ministry of Education No.
NRF-2022R1I1A1A01068833.
We thank APCTP, Pohang, Korea for their hospitality during the Focus Program [APCTP-2026-T04] from which this work greatly benefited.

%\begin{appendices}
%\renewcommand{\thesection}{\Alph{section}}
%\numberwithin{equation}{section}

%\section{Zero temperature analysis of D7 brane}
%\label{sec:appen_A}
%% When $T=0$, $\xi_H =0$,
%\end{appendices}

%%%%%%%%%%%%%%%%%%%%%%%%%%%%%%%%%%%%%%%%%%%%%%%%%%
%%%%%%%%%%%%%%%%%%%%%%%%%%%%%%%%%%%%%%%%%%%%%%%%%%
\bibliographystyle{JHEP}
\bibliography{Weyl_inst}
%%%%%%%%%%%%%%%%%%%%%%%%%%%%%%%%%%%%%%%%%%%%%%%%%%
%%%%%%%%%%%%%%%%%%%%%%%%%%%%%%%%%%%%%%%%%%%%%%%%%%

%%%%%%%%%%%%%%%%%%%%%%%%%%%%%%%%%%%%%%%%%%%%%%
%\begin{thebibliography}{99}

%%%%%%%%%%%%%%%%%%%%%%%%%%%%%

%\end{thebibliography}

\end{document}